\documentclass[fleqn,usenatbib]{mnras}
\usepackage{amsmath}	
\usepackage{txfonts}

\usepackage[T1]{fontenc}

\DeclareRobustCommand{\VAN}[3]{#2}
\let\VANthebibliography\thebibliography
\def\thebibliography{\DeclareRobustCommand{\VAN}[3]{##3}\VANthebibliography}

\usepackage{graphicx}	
\usepackage{mhchem}	
\usepackage{xcolor}	
\usepackage{multirow}
\usepackage{footnote}
\makesavenoteenv{tabular}
\makesavenoteenv{table}
\usepackage{lipsum}

\title[Hydroxylamine Chemistry in the ISM]{Processing of hydroxylamine, \ce{NH2OH}, an important prebiotic precursor, on interstellar ices.}

\author[G. Molpeceres et al.]{
Germ\'an Molpeceres,$^{1}$\thanks{E-mail: molpeceres@astron.s.u-tokyo.ac.jp}
V\'ictor M. Rivilla,$^{2}$
Kenji Furuya,$^{3}$
Johannes K\"astner,$^{4}$
Bel\'en Mat\'e,$^{5}$
Yuri Aikawa.$^{1}$
\\
$^{1}$Department of Astronomy, Graduate School of Science, The University of Tokyo, 113-0033, Tokyo, Japan\\
$^{2}$Centro de Astrobiología (CSIC, INTA), Ctra. de Ajalvir, km. 4, Torrejón de Ardoz, E-28850 Madrid, Spain\\
$^{3}$National Astronomical Observatory of Japan, 181-8588, Tokyo, Japan. \\
$^{4}$Institute for Theoretical Chemistry, University of Stuttgart, Pfaffenwaldring 55, 70569 Stuttgart, Germany \\
$^{5}$Instituto de Estructura de la Materia (IEM-CSIC), Calle Serrano 113, 28006, Madrid, Spain.
}

\date{Accepted XXX. Received YYY; in original form ZZZ}

\pubyear{2015}

\begin{document}
\label{firstpage}
\pagerange{\pageref{firstpage}--\pageref{lastpage}}
\maketitle

\begin{abstract}
Hydroxylamine, \ce{NH2OH}, is one of the already detected interstellar molecules with the highest prebiotic potential. Yet, the abundance of this molecule found by astronomical observations is rather low for a relatively simple molecule, $\sim$ 10$^{-10}$ relative to H2. This seemingly low abundance can be rationalized by destruction routes operating on interstellar dust grains. In this work, we tested the viability of this hypothesis under several prisms, finding that the origin of a lower abundance of \ce{NH2OH} can be explained by two chemical processes, one operating at low temperature (10 K) and the other at intermediate temperature (20 K). At low temperatures, enabling the hydrogen abstraction reaction \ce{HNO + H -> NO + H2}, even in small amounts, partially inhibits the formation of \ce{NH2OH} through successive hydrogenation of NO, and reduces its abundance on the grains. We found that enabling a 15--30\% of binding sites for this reaction results in reductions of \ce{NH2OH} abundance of $\sim$ 1-2 orders of magnitude. At warmer temperatures (20 K, in our study), the reaction \ce{NH2OH + H -> HNOH + H2}, which was found to be fast (k$\sim$10$^{6}$ s$^{-1}$) in this work, followed by further abstractions by adsorbates that are immobile at 10 K (O, N) are the main route of \ce{NH2OH} destruction. Our results shed light on the abundance of hydroxylamine in space and pave the way to constraining the subsequent chemistry experienced by this molecule and its derivatives in the interstellar prebiotic chemistry canvas. 
\end{abstract}




\section{Introduction}

Understanding the origin of life is one of the so-called ``Holy Grails of Chemistry" \citep{burrows2017}. To take steps towards this ambitious goal, one of the most fundamental conundrums remains to establish the transition from purely abiotic compounds to prebiotic ones (i.e aminoacids, nucleotides, lipids and sugars) \citep{plankensteiner2004,Chandru2016,Yadav2020}. Postulating routes of formation of prebiotic molecules is a crucial first step for more advanced origin of life theories, such as the ``RNA World Hypothesis" \citep{Gilbert1986}, one of the most accepted theories for the emergence of life, which proposes that ribonucleic acid (RNA) may have played a twofold role in primordial Earth, first as a carrier of genetic material and second as a catalyst in the biosynthesis of proteins and the subsequent DNA/RNA/Proteins cycle.

Many of the prebiotic precursors do not present a complicated chemical structure and, consequentially, are susceptible to being detected in the interstellar medium (ISM). For example, recent astronomical surveys have found 1-2-ethenediol  (\ce{(C2H2O)2}\citep{rivilla2022ethenediol} a key molecule in the synthesis of sugars, ethanolamine (\ce{NH2CH2CH2OH}) \citep{Rivilla2021} a key molecule in the synthesis of phospholipids, and the molecule subject of this study, hydroxylamine \ce{NH2OH} \citep{Rivilla_2020}, which is assumed to be crucial in forming amino acids and RNA nucleotides \citep{sakurai_prebiotic_1984,snowGasPhaseIonicSyntheses2007,becker2019, xu_isoxazole_2022}.

The presence of these prebiotic precursors in space is a fascinating subject that opens many branches, such as the following. How are these molecules synthesized under interstellar conditions? How do they evolve? Are they the summit of complexity expectable in the ISM? These questions require tailored, experimental or theoretical studies to reveal these precursors' physical and chemical nature under interstellar conditions. For \ce{NH2OH}, several of these questions have been explicitly addressed. For example, literature shows that it is produced mostly through nitric oxide (\ce{NO}) hydrogenation \citep{Congiu2012, fedoseev2012, fedoseev_simultaneous_2016, nguyen_experimental_2019} on the surface of dust grains. Alternative routes include non-energetic oxidation of ammonia \citep{He2015b}, e.g.  by cosmic ray irradiation of \ce{O2} and \ce{NH3} ices \citep{tsegawFormationHydroxylamineLowTemperature2017}. Finally, the classical route for hydroxylamine is the radical recombination of \ce{NH2} and OH \citep{nishi_photodetachment_1984}.  Concerning the evolution of hydroxylamine, a significant theoretical and experimental effort has been dedicated to determining its reactivity in the gas phase, primarily through its ionized (\ce{NH2OH+}) form \citep{snowGasPhaseIonicSyntheses2007, Largo2009, Barrientos2012, Redondo2013b, SanzNovo2019}. 

One key issue surrounding hydroxylamine prevalence under interstellar conditions is its elusive nature. For a relatively simple molecule containing five atoms, with equally standard functional groups, e.g. amino \ce{-NH2} and hydroxyl \ce{-OH}, models predict total abundances in the range between 10$^{-5}$--10$^{-6}$ \citep{Garrod2013, He2015b, Garrod2022}. However, the abundance of \ce{NH2OH} in the ISM is surprisingly low, of 2.1 $\times$ 10$^{-10}$ to \ce{H2} \citep{Rivilla_2020}. Besides, \ce{NH2OH} has only been detected towards our galactic centre, specifically in the molecular cloud G+0.693-0.027, an astronomical environment characterized by rather unique conditions in terms of energetic input. For example, temperatures of the gas in the range between 50 and 150 K \citep{Krieger2017, Zeng2018}, dust temperatures of 15 K \citep{rodriguez-fernandez2004}, and most importantly, are subjected to mechanical shocks due to cloud-cloud collisions \citep{martin-pintado1997} and a high cosmic-ray ionization rate ($\sim 10^{-15}-10^{-14}$ s$^{-1}$) \citep{Zeng2018, goto2013} that might alter the molecular inventory of this region. Hydroxylamine has not been detected in other regions, neither in cold environments (molecular clouds) nor warmer ones (hot-corinos and hot-cores) \citep{pulliam_search_2012, mcguire_cso_2015, codella_nitrogen_2018, ligterink_alma-pils_2018, }. The upper limits for Sgr B2 and IRAS 16293-2422 are 8$\times$10$^{-12}$ and 3.1$\times$10$^{-11}$, respectively.

The deficiency of \ce{NH2OH} could be due to \ce{NH2OH} being liable for being chemically converted (destroyed) on the surface of interstellar dust grains through hydrogenations and other reactions \citep{Garrod2013, Garrod2022}. In this study, we have explicitly addressed this scenario under a series of possible reactions, namely isomerization reactions (cis-trans, c-\ce{NH2OH}/t-\ce{NH2OH}), irreversible destructions to produce the highly stable \ce{H2O} and \ce{NH3} molecules, and reversible destructions through an H-addition and H-abstraction cycle \citep{Molpeceres2022}. We study these reactions using high-precision quantum chemical calculations. In particular, the reactions that we study are:

\begin{align}
    \ce{c-NH2OH &<=> t-NH2OH} \\
    \ce{NH2OH + H &-> NH3 + OH} \label{eq:res1} \\
    \ce{NH2OH + H &-> H2O + NH2} \label{eq:res2} \\
    \ce{NH2OH + H &-> H2NO + H2} \label{eq:res3} \\
    \ce{NH2OH + H &-> HNOH + H2} \label{eq:res4} \\
    \ce{HNOH + H &-> NH2OH} \label{eq:brok:1} \\
    \ce{H2NO + H &-> NH2OH } \label{eq:brok:2}.
\end{align}

Later, we merge the results attained in the present work with further works available in the literature \citep{He2015b, nguyen_experimental_2019} and complement our theoretical calculations with tailored astrochemical models. This article's structure is as follows. In section \ref{sec:methods} we present our computational framework for studying the candidate reactions for the destruction of \ce{NH2OH}. In section \ref{sec:results}, we give a complete picture of the destruction and reformation of \ce{NH2OH} through isomerization and hydrogenation, in combination with previous works \citep{nguyen_experimental_2019}. Section \ref{sec:discuss} builds on the quantum chemical findings and provides a comprehensive analysis of the importance of these reactions in the bigger astrochemical picture. Also, we reanalyze some of the previously thought formation reactions of \ce{NH2OH}. Then we include reactions in the literature and our data into astrochemical models. Finally, we provide an overview and outlook in Section \ref{sec:conclusion}.

\section{Methods} \label{sec:methods}

\subsection{Quantum Chemical Calculations}

\subsubsection{Reactivity} \label{sec:met:react}

The study of the reactivity of hydroxylamine uses density functional theory (DFT) in combination with high-level calculation to refine the electronic energy. In particular, geometries and vibrational frequencies were obtained using the DSD-PBEP86-D3BJ double hybrid exchange and correlation functional \citep{Kozuch2011,Grimme2011}, and the def2-TZVP basis set \citep{Weigend2005}. The electronic energies were refined utilizing DLPNO-CCSD(T) (domain-based local pair natural orbital coupled cluster) calculations using a two-point extrapolation to the basis set limit (CBS) \citep{Helgaker1997, Zhong2008, Neese2011, guo_communication_2018}. For the extrapolation, we used the energies obtained using with the cc-pVTZ, and cc-pVQZ basis sets \citep{Woon1993}. The level of theory used in the reactivity calculations can be therefore abbreviated as DLPNO-CCSD(T)/CBS//DSD-PBEP86-D3BJ/def2-TZVP. The frozen-core approximation was applied in all wavefunction theory-based calculations. The DFT and DLPNO-CCSD(T) calculations include resolution of the identity techniques to speed up the calculations \emph{via} the RIJK approximation for the Coulomb and Exchange integrals additional RI approximations for the correlation part of the exchange and correlation evaluation. 

We used two different structural models for all the reactions we present in Section \ref{sec:results}. In the first place (Model \textbf{A}), we used a gas-phase structural model, and then the derived rate constants were corrected to account for the fixated rotational degrees of freedom by fixing the rotational partition function on the surface to the unity. Such an approach, named the implicit surface approach, is helpful in many cases where the surface does not play an active role in the reaction (\cite{Molpeceres2021b,Molpeceres2022b} show two recent examples of such approach). 

\begin{figure}
    \centering
    \hspace{0.10cm}
    \includegraphics[width=\linewidth]{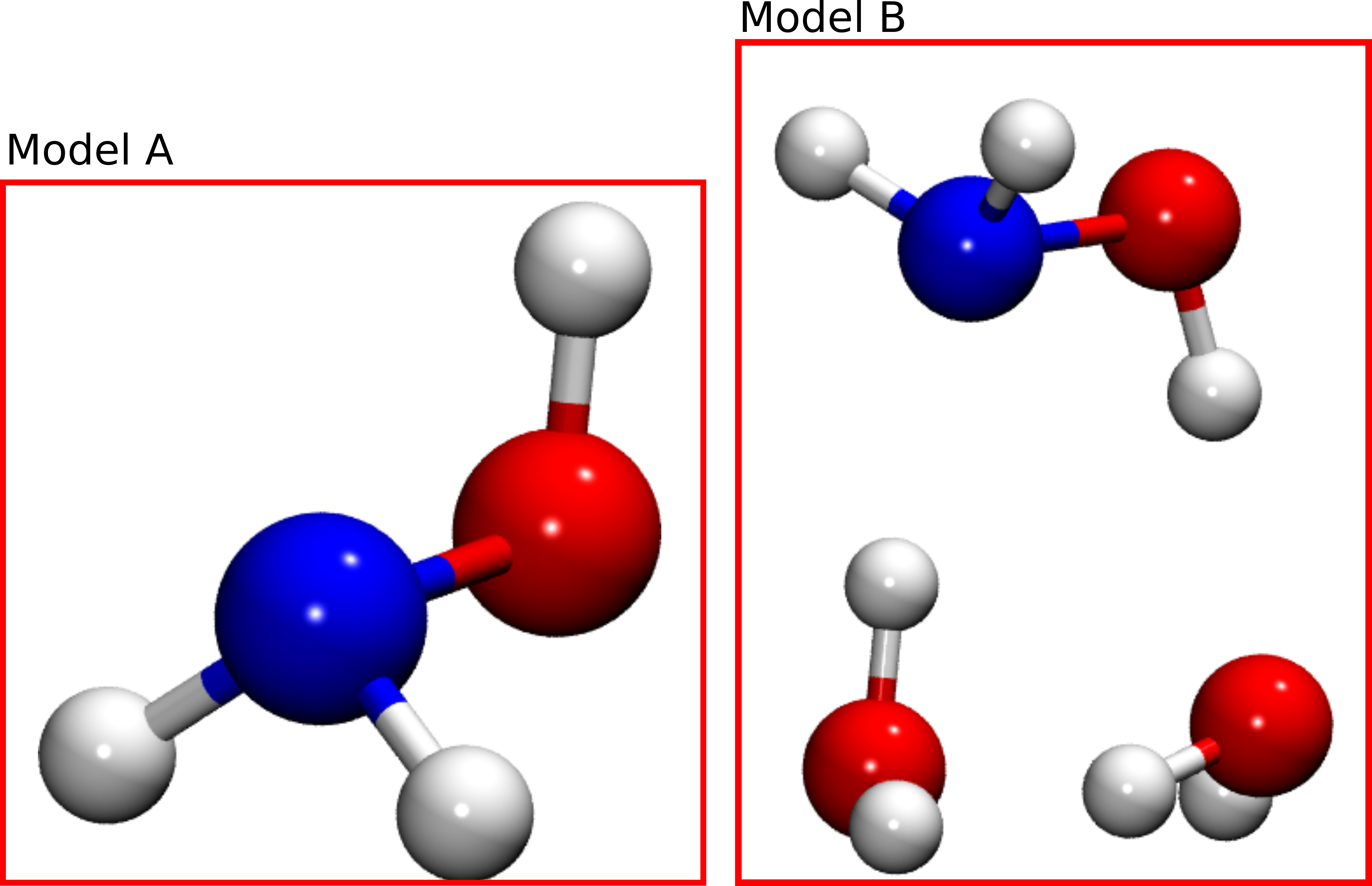}
    \caption{Structural models employed for the study of the reactivity of hydroxylamine (left) Implicit Surface (Model \textbf{A}). (right) Two model water approach (Model \textbf{B}). The donor-acceptor nature of \ce{NH2OH} can be inferred from the figure.}
    \label{fig:GasH2OModel}
\end{figure}

Hydroxylamine, as already pointed out in \cite{Wakelam2017}, has a peculiar behaviour when interacting with a water surface because of its concomitant behaviour as an H-donor and H-acceptor in hydrogen bonds. Hence, we also used a model (Model \textbf{B}) explicitly considering two water molecules but maintaining the fixing of the rotational partition function to account for possible orientation effects stemming from the -\ce{NH2} and -\ce{OH} interaction with water. The initial configuration for Model \textbf{B} is obtained from exploratory calculations of the interaction of \ce{NH2OH} with a water dimer in its global minimum. A total of 25 initial positions, for \ce{NH2OH} were tried, from where the configuration with the lowest energy is used for further calculations. An example of the two different models for the adsorption of \ce{NH2OH} is depicted in Figure \ref{fig:GasH2OModel}.

The formalism for studying the reactions is based on transition state theory (TST) incorporating quantum nuclear effects from semiclassical instanton theory \citep{lan67,mil75,col77,Rommel2011-2, Rommel2011}. Thus, for a given reaction, we determine the stationary points in the potential energy surface (reactant, transition states, and products) from where we extract the classical (e.g. not including quantum effects) rate constants. These are later expanded by locating the instantons at different temperatures, e.g. performing a sequential cooling scheme. The discrete nature of the rate constants obtained by instanton theory, in contrast to the classical continuous counterparts, makes the calculation of these quantities and their convergence behaviour at low temperatures expensive and complicated. For all the reactions presented in Section \ref{sec:results} we indicate the lowest temperature of our calculation, assuming that the extrapolation to lower temperatures should yield very similar values granted by the asymptotic behaviour of tunnelling-dominated rate constants at low temperatures. The magnitude that controls the temperature at which we start to include quantum effects (e.g. tunnelling starts to play a significant role) is the crossover temperature:

\begin{equation}
    T_\text{c} = \frac{\hbar \nu_{i}  }{k_\text{B}},
\end{equation}

\noindent where $\nu_{i}$ is the transition frequency at the transition state (e.g. imaginary frequency). Additionally, since our work aims to refine astrochemical models, we have not included symmetry factors ($\sigma$) in our rate constants. Even if in the gas phase, they may be present, on interstellar ice, the rotational symmetry is broken. Similar to the PES exploration, our instanton calculations employ a dual-level formalism \citep{Meisner2018}.

Finally, we considered radical-radical reactions in the reformation of hydroxylamine through reactions \ref{eq:brok:1} and \ref{eq:brok:2}.
These two reactions are the only ones relevant to the formation of hydroxylamine after the reaction of \ce{NH2OH} with an H atom; see Section \ref{sec:results:destruction}. They were simulated using a broken-symmetry formalism, first converging the pre-reactant complex's high-spin state, followed by downhill optimizations using the correct spin multiplicity. This set of calculations requires a more extensive sampling, and therefore we reduced the basis set of the DSD-PBEP86-D3BJ calculations to def2-SVP \citep{Weigend2005}. We initially sampled 150 starting positions for the downhill optimizations, from which we ensured that less than 10\% did not converge, a sometimes unavoidable outcome due to the complicated electronic structure of the system. All the calculations use the ORCA (v.5.0.2) software \citep{Neese2020} interfaced with Dl-Find/Chemshell \citep{Chemshell, kae09a}. The sampling of reaction outcomes for the radical-radical reactions follows the procedure presented in Section \ref{sec:met:binding} for Model \textbf{A}, modified to account for the presence of water molecules in Model \textbf{B} (see Method section in \citealt{Molpeceres2021b}).

\subsubsection{Binding energy calculations} \label{sec:met:binding}

Binding energies are critical in grain surface chemistry, interpreted as the attractive interaction between an adsorbate and a surface. To account for a diversity of binding sites stemming from the amorphous solid water (ASW) structure, we have sampled the distribution of binding energies of the primary intermediates of the reactions considered in this work (e.g. \ce{NH2OH}, \ce{HNOH}, and \ce{H2NO}), and incorporated these values into the astrochemical model. The structural models for the ices are the same \ce{(H2O)20} clusters that were used in \citet{Molpeceres2021b}. These clusters were constructed following a heating and cooling cycle of molecular dynamics simulations with an empirical force field followed by a geometry optimization using an electronic structure method. Once the clusters are prepared, the above-mentioned molecules are situated at a distance given by a tolerance parameter plus the maximum \{$x$, $y$, $z$\} value of the cluster. The orientation of the molecules is randomized, and the system is allowed to relax \emph{via} a geometry optimization. The binding energies are then obtained as the difference between the bimolecular system of cluster and adsorbate and the complex formed by the relaxed molecule on the cluster. We sampled ten initial positions per cluster for five clusters (50 binding energies per considered molecule). More details on the protocol can be found in \cite{Molpeceres2021P}. For computational saving reasons and due to the considerable binding energy found for the adsorbates, we omitted ZPVE in calculating the binding energies, and we fixed the cartesian positions of all the water molecules 8\AA{} away from any of the initial positions of the adsorbate. While ZPVE effects are known to shift the binding energy distributions, the high binding energy from the electronic energy contribution is high enough to assume that this effect will be minor at 10--20 K.

The electronic structure method we used for reactivity, presented in Section \ref{sec:met:react} is too expensive to sample binding energies. For the binding energy calculations, we used the B97-3c method in the geometry optimizations \citep{brandenburgB973c} with energy refinement coming from the DLPNO version of the DSD-PBEP86-D3BJ. Hence, the method is abbreviated as DLPNO-DSD-PBEP86-D3BJ/def2-TZVP//B97-3c. It is worth mentioning that even though the level of theory for the binding energies is more approximated than the local coupled cluster methods used in the reactivity part of this work, using a double-hybrid DFT method for the electronic energies should ensure good treatment of the weak interatomic interactions in the adsorbate-ice system \citep{Kozuch2011}.

\section{Quantum Chemical Results} \label{sec:results}

\subsection{Isomerism of Hydroxylamine on ASW} \label{sec:isomerism}

The first reaction we looked at is the interconversion in the gas and on the grains of trans-\ce{NH2OH} (t-\ce{NH2OH}) and cis-\ce{NH2OH} (c-\ce{NH2OH}). This is because we evaluated several hypotheses for the mismatch between observations \citep{Rivilla_2020} and models \citep{Garrod2013, Garrod2022}, and one of them could be preferential destruction of a particular isomer on the grain or isomerization in the gas. So far, only t-\ce{NH2OH} has been detected in the ISM\citep{Rivilla_2020}. Both isomers are separated by $\sim$ 2,300 K in the gas phase at the DLPNO-CCSD(T)/CBS//DSD-PBEP86-D3BJ/def2-TZVP level of theory. \ce{NH2OH} can isomerize via two distinct molecular motions. First, \emph{via} nitrogen inversion, e.g. simultaneous migration of the two hydrogen atoms for the \ce{NH2} moiety. Second, \emph{via} torsion of the HONH bond. While the former is potentially affected by quantum tunnelling, this effect on the latter is minor. 

We sampled both molecular motions for Model \textbf{A} and Model \textbf{B}. In this case, for Model \textbf{A}, we do not fix the rotational partition function because we are interested in the gas-phase process of inter-converting trans and cis isomers. The energetic quantities describing the process are gathered in Table \ref{tab:isomerization}. We attribute an increase in the activation energies between Model \textbf{A} and Model \textbf{B} to restraining motions on the surface. In the case of the isomerization \emph{via} the torsion motion, this increase is not very pronounced ($\pm$ 200 K). Still, for the inversion of nitrogen, the activation energy change is $\pm$ 1600 K. The reaction energies are similar both in Model \textbf{A} and Model \textbf{B}, indicating a similar binding of cis and trans isomers to the surface. Finally, for the imaginary transition mode, we observe minor differences between both mechanisms. However, for the torsion mechanism, a reduction of merely 200 cm$^{-1}$ in the transition frequencies significantly affects the rate constants, as we observed for the OCSH radical on ice surfaces \citep{Molpeceres2021b}.

\begin{table}
    \caption{Molecular mode (Nitrogen Inversion or Torsion of HONH bond), reaction energies ($\Delta$U$_\text{R}$ in K), activation energies ($\Delta$U$_\text{a}$, in K) and transition frequencies ($\nu_{i}$ in cm$^{-1}$) for the \ce{c-NH2OH -> t-NH2OH} reaction. The direction of the reaction is chosen for the reaction to be exothermic. }
    \label{tab:isomerization}
    \centering
    \begin{tabular}{cccccc}
    \hline
        Structural Model & Mode & $\Delta U_{R} (K)$ & $\Delta U_{a}$ (K) & $\nu_{i}$ (cm$^{-1}$) \\
    \hline
    Model A (Cis$\rightarrow$Trans) & Inversion & -2,026 & 3,666 & 956$i$\\
    Model B (Cis$\rightarrow$Trans) & Inversion & -2,291 & 5,274 & 865$i$\\   
    Model A (Cis$\rightarrow$Trans) & Torsion   & -2,026 & 1,091 & 476$i$\\
    Model B (Cis$\rightarrow$Trans) & Torsion   & -2,291 & 1,257 & 257$i$\\
    \hline
    \end{tabular}
\end{table}

\begin{figure*}
    \centering
    \includegraphics[width=8cm]{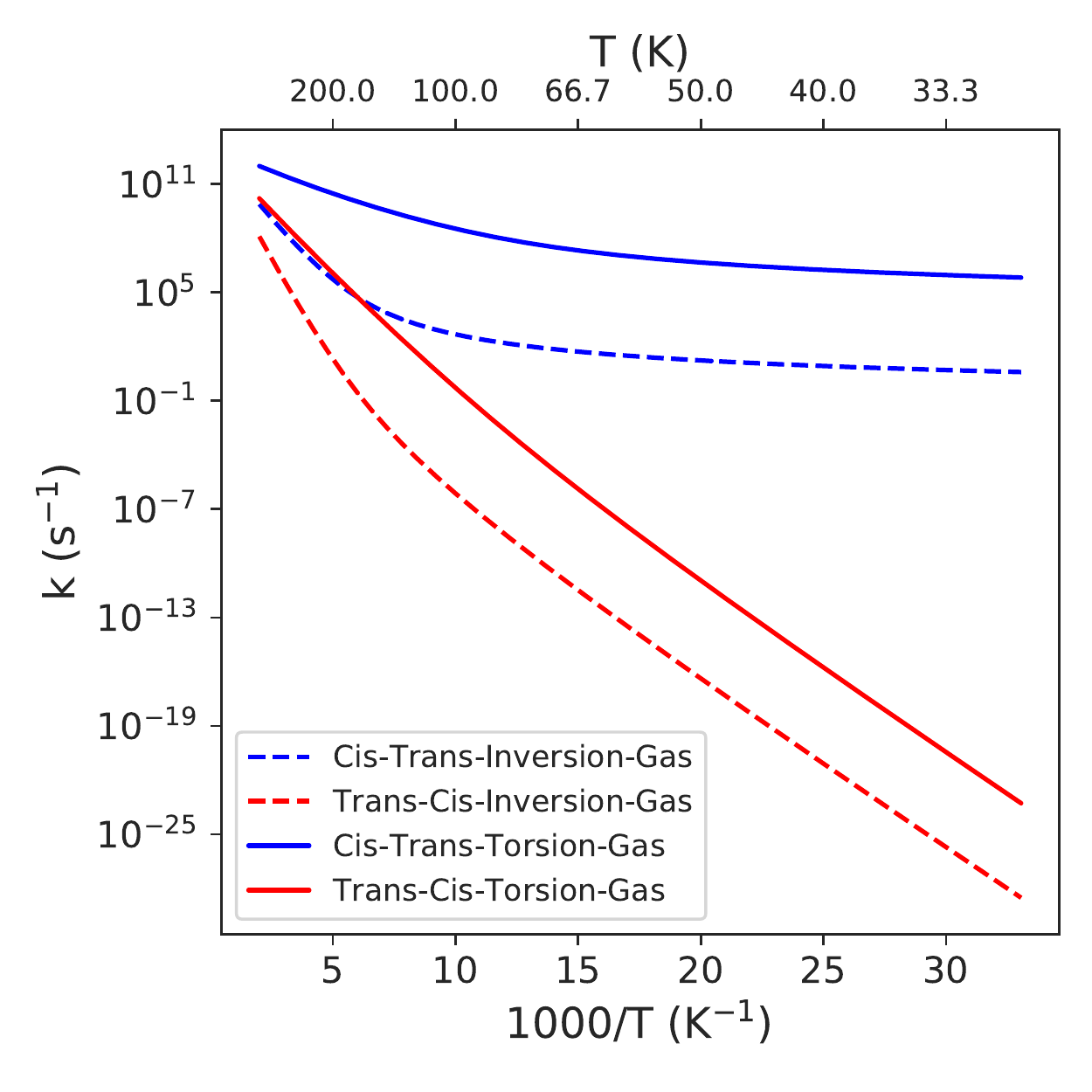}
    \includegraphics[width=8cm]{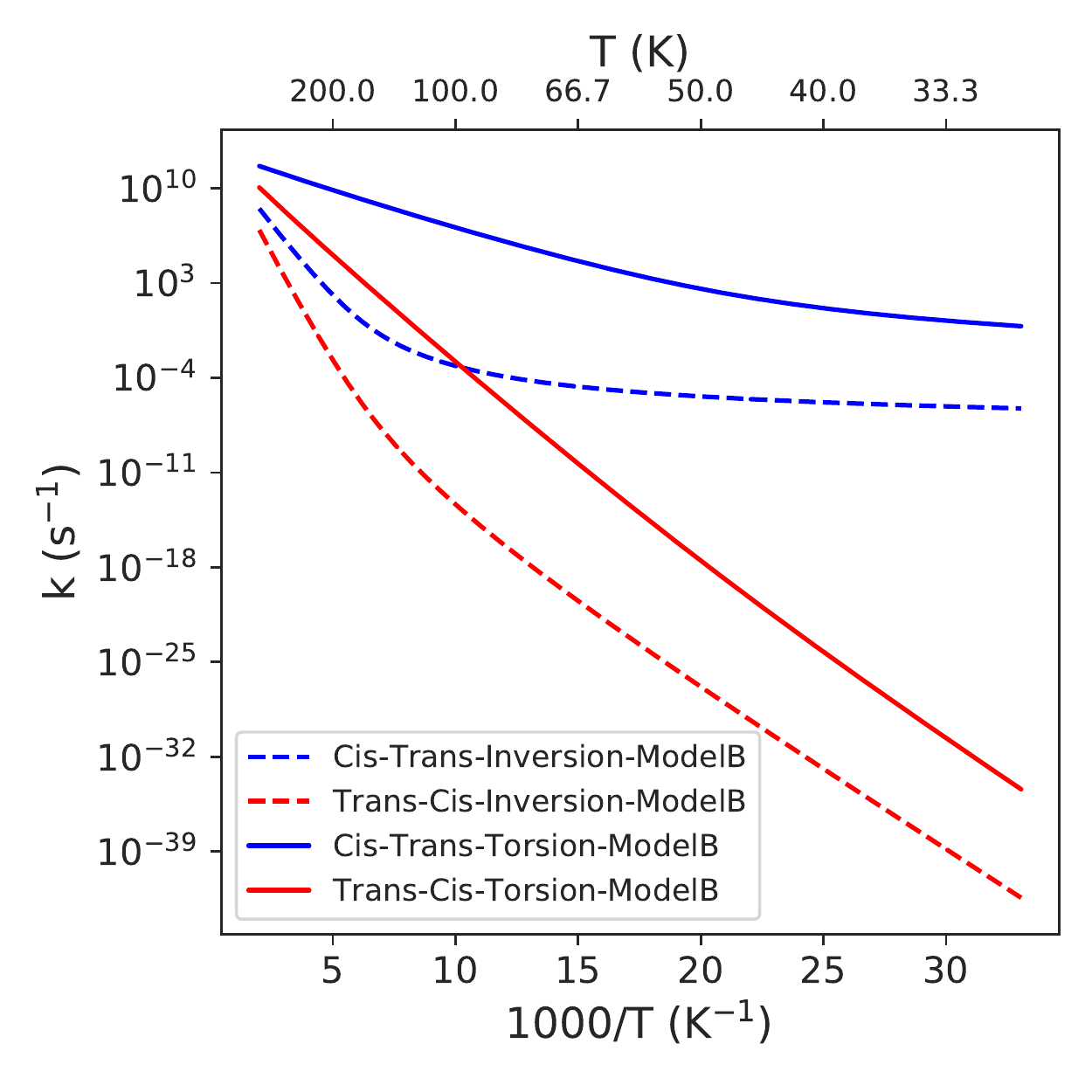}
    \caption{Rate constants for isomerizations in \ce{NH2OH}. Left panel - Isomerization rate constants in the gas. Right panel - Isomerization rate constants in the 2\ce{H2O} explicit model (Model \textbf{ B}) }
    \label{fig:rates_isomerization}
\end{figure*}

The endothermicity of the Trans$\rightarrow$Cis reaction, summed to the high activation energies of the inversion reaction (more affected by tunnelling), make that, for this particular reaction, we calculated the rate constants of reaction employing a transition state theory corrected by a tunnelling factor obtained from the fit of an asymmetric Eckart barrier. The rate constants are presented in Figure \ref{fig:rates_isomerization}. From the trends, we observe that, in general, rate constants in Model \textbf{B} are lower than in the gas phase (Model \textbf{A}). This is a consequence of the steric hindrance produced by the water molecules. However, in both cases, cis $\rightarrow$ trans isomerization is a fast process in the gas and the surface. All rate constants for trans-cis isomerization are negligible at low temperatures, indicating that any t-\ce{NH2OH} should not convert to c-\ce{NH2OH}. Likewise, any c-\ce{NH2OH} formed through chemical reactions (see, for example, Section \ref{sec:brokensym}) must isomerize quite fast under astrophysical conditions (e.g. below 150 K). Even the lowest rate constant for \ce{c-NH2OH -> t-NH2OH} is on the order of 10$^{-4}$ s$^{-1}$ that is higher than the estimated accretion rate of H of 10$^{-5}$ s$^{-1}$ \citep{Wakelam2017H2}, which translates in timescales of 10$^{4}$ and 10$^{5}$ s$^{-1}$, respectively. Hence, in the following, any chemistry of \ce{NH2OH} will imply t-\ce{NH2OH}, assuming that abundances of c-\ce{NH2OH} must be negligible. Finally, it is interesting to mention that \ce{NH2OH} isomerization through the classical torsion barrier is faster for the whole considered temperature range than through nitrogen inversion, even considering that tunnelling is more critical for the latter. This is due to the significantly lower value of the torsional barrier.

\subsection{Destruction of \ce{NH2OH}} \label{sec:results:destruction}

In addition to the reactions presented in \citet{nguyen_experimental_2019} and \citet{He2015b}, whose implications are discussed in Section \ref{sec:discuss}, we have studied reactions \ref{eq:res1} to \ref{eq:brok:2} in  Model \textbf{A} and Model \textbf{B}. The rationale behind the explicit study of these reactions is to determine viable \ce{NH2OH} destruction routes to understand the abundances observed in \cite{Rivilla_2020}. A summary of the energetic descriptors for the reaction can be found in Table \ref{tab:energetics} for a quick overview of the results. At the same time, in the text, we deepen in the discussion of the reactions from a chemical point of view. 

\begin{table}
    \caption{Reaction energies, $\Delta U_{R}$, and activation energies, $\Delta U_{a}$ (in K, including ZPVE), for the reactions proceeding with an activation barrier, and studied using quantum chemical methods in the present work.}
    \label{tab:energetics}
    \centering
    \begin{tabular}{ccccccc}
    \hline
    Reaction & Label &Model & $\Delta U_{R}$ (K) & $\Delta U_{a}$ (K) \\
    \hline
    \multirow{2}{*}{\ce{NH2OH + H -> NH3 + OH}} & \ref{eq:res1}  & \textbf{A} & -24,893 & 3,723 \\
                                                & \ref{eq:res1}  & \textbf{B} & -24,695 & 4,903 \\
    \multirow{2}{*}{\ce{NH2OH + H -> H2O + NH2}}& \ref{eq:res2} & \textbf{A} & -29,762 & 6,437 \\
                                                & \ref{eq:res2} & \textbf{B} & -24,877 & 6,206 \\
    \multirow{2}{*}{\ce{NH2OH + H -> H2NO + H2}}& \ref{eq:res3} & \textbf{A} & -13,172 & 4,206 \\
                                                & \ref{eq:res3} & \textbf{B} & -13,444 & 7,439 \\
    \multirow{2}{*}{\ce{NH2OH + H -> HNOH + H2}}& \ref{eq:res4} & \textbf{A} &  -9,527 & 3,409 \\
                                                & \ref{eq:res4} & \textbf{B} & -10,763 & 3,218 \\                   
    \hline
    \end{tabular}
\end{table}

\subsubsection{\ce{NH2OH + H -> NH3 + OH}} \label{sec:NH3}

The first of the destruction reactions that we studied is the addition of a hydrogen atom to the \ce{NH2} moiety of hydroxylamine, leading to ammonia and the hydroxyl radical. Subsequently, the reaction:

\begin{equation}
    \ce{NH3 + OH -> H2O + NH2}\label{eq:h2o_nh2},
\end{equation}

\noindent is supposed to dominate at warm temperatures (k=1.6$\times$10$^{-13}$ cm$^{3}$s$^{-1}$ at 300 K) according to the KIDA database and references therein \citep{wakelam_kinetic_2012,atkinson_evaluated_2004,Sander2011}. It is important to note that reaction \eqref{eq:h2o_nh2} is only tested for intermediate temperatures (230--450 K) and in the gas phase. Hence low temperature surface studies are desirable. Therefore, reaction \eqref{eq:res1} would result in a total loss of hydroxylamine and a reduction of the abundance of this molecule in the ISM. 

\begin{figure}
    \centering
    \hspace{0.10cm}
    \includegraphics[width=4cm]{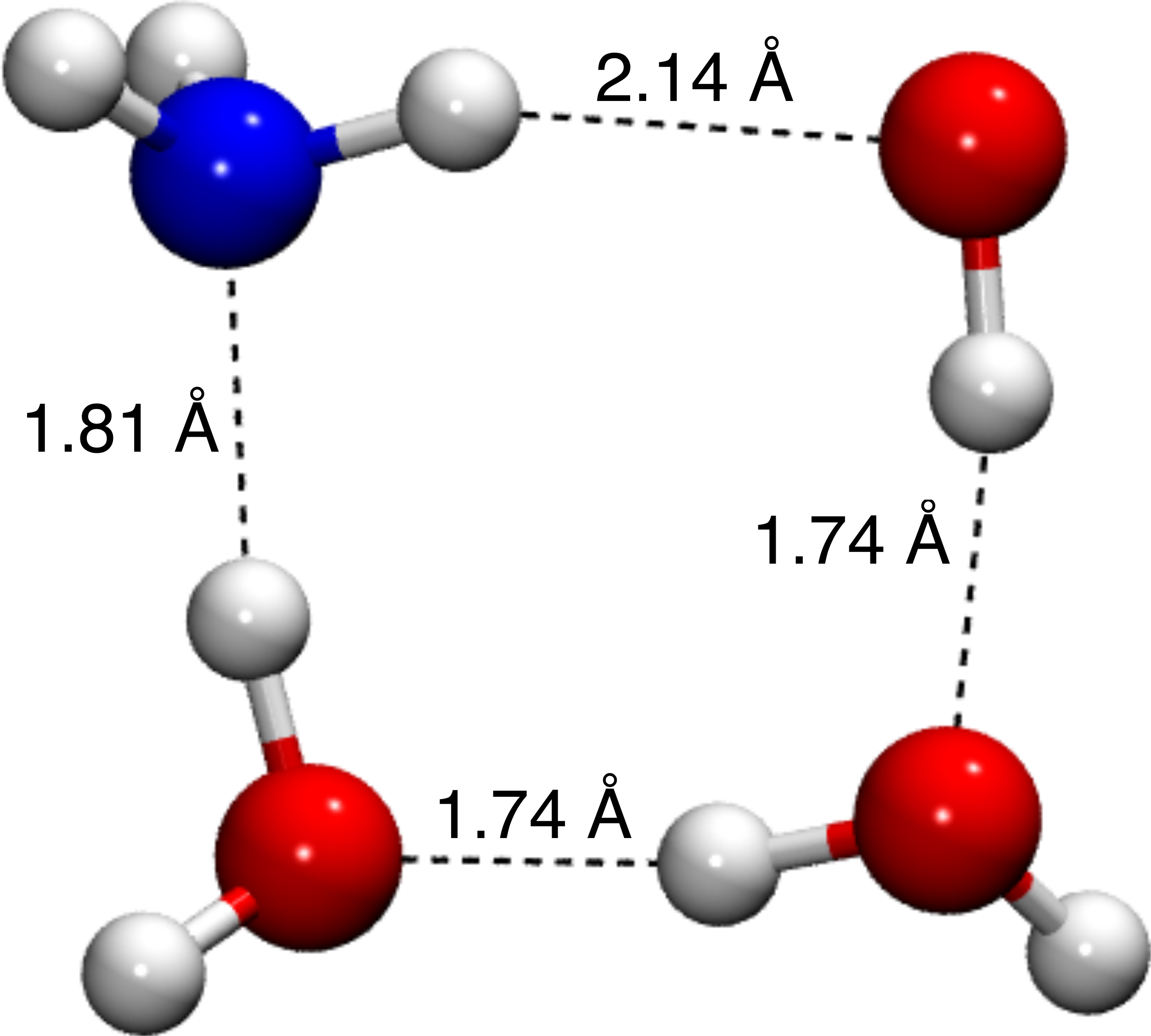} \vline
    \hspace{0.10cm}
    \includegraphics[width=4cm]{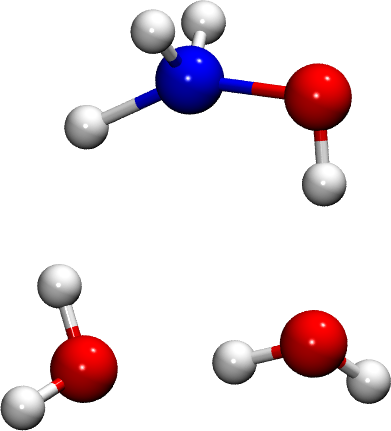} 
    \caption{(Left). Product of reaction \eqref{eq:res1} in Model \textbf{B}. As a reference, the OH-H distance in the water dimer is 1.94~\AA~at the DLPNO-CCSD(T)/CBS//DSD-PBEP86-D3BJ/def2-TZVP level of theory. (Right). Transition state for reaction \ref{eq:res1} in Model \textbf{B}.}
    \label{fig:NH3_OH}
\end{figure}

Reaction \ref{eq:res1} is exothermic both in Model \textbf{A} and Model \textbf{B}, with energies of reaction of $\Delta U_{R, \text{ModelA}}$=-24,893 K and $\Delta U_{R, \text{ModelB}}$= -24,695 K, e.g. a tiny influence of the addition of explicit water molecules to the reaction. The high reaction energy can be explained by the strong interaction of the \ce{NH3-OH} pair formed during the reaction. This is not the case for activation energies, with values: $\Delta U_{a, \text{ModelA}}$=3,723 K and $\Delta U_{a, \text{ModelB}}$=4,903 K. While normally, and especially for weakly bound adsorbates, the ASW surface does not play a significant role \citep{Meisner2017, Lamberts2017, Molpeceres2022b} the case of hydroxylamine seems particular, owing to the donor-acceptor nature of the \ce{NH2OH-2H2O} complex. The reaction's increased exothermicity and higher activation energy in Model \textbf{B} can be visualized in Figure \ref{fig:NH3_OH}. Contributing to the reaction exothermicity, a cooperative four-centre network is formed in the product, and contributing to the high activation energy we find a temporary break of a hydrogen bond to initiate the reaction.

\begin{figure}
    \centering
    \includegraphics[width=\linewidth]{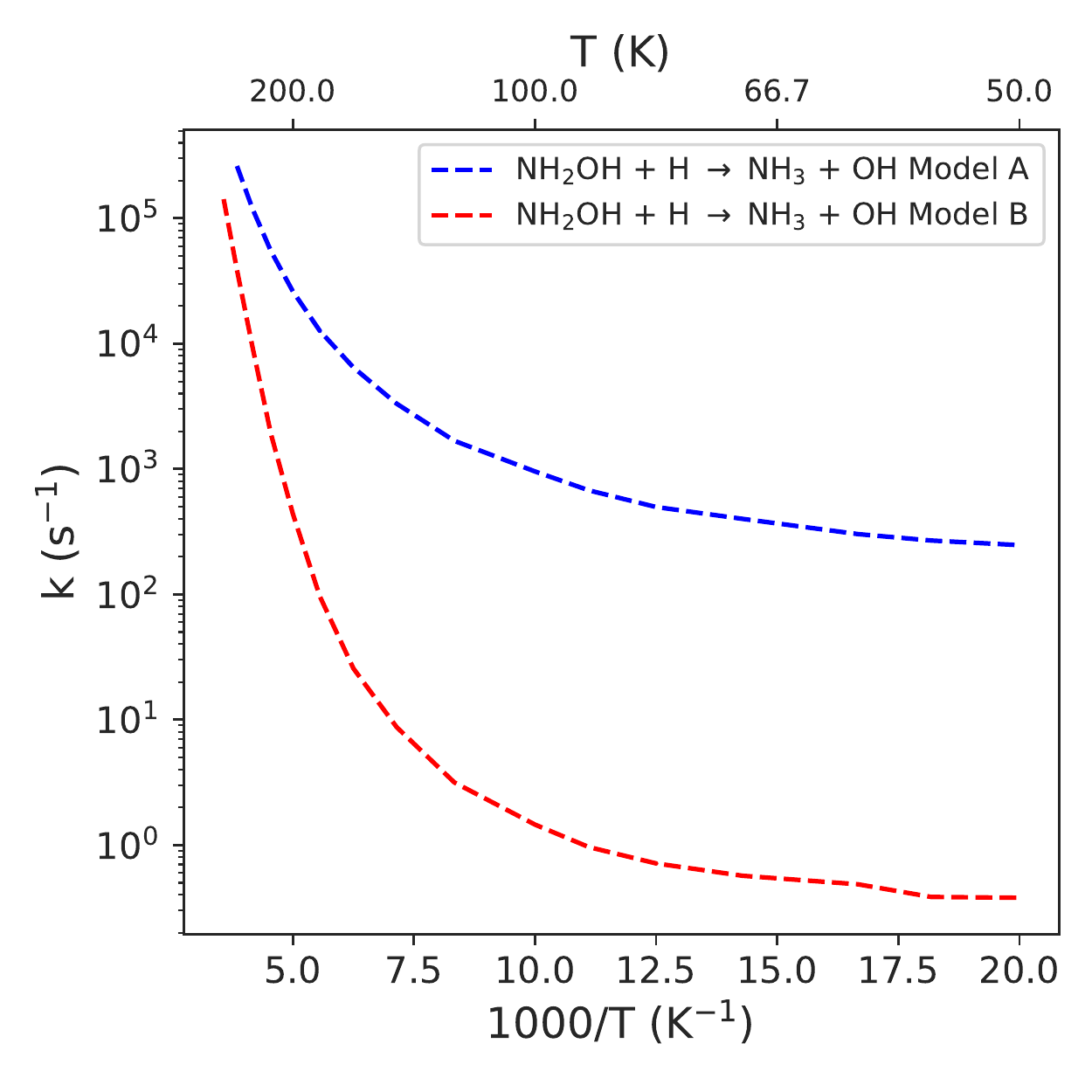}
    \caption{Instanton rate constants for reaction \eqref{eq:res1} in Model \textbf{A} and Model \textbf{B}. The crossover temperature is 300 K for the former and 308 K for the latter.}
    \label{fig:rates_nh3}
\end{figure}

Instanton rate constants for reaction \eqref{eq:h2o_nh2} down to 50 K are presented in Figure \ref{fig:rates_nh3}, a temperature where the asymptotic regime for tunnelling is reached. We note a decrease in the reaction rate constants of 3 orders of magnitude in the case of Model \textbf{B}, which is a better descriptor of deep binding sites, compared with Model \textbf{A}. Nonetheless, this reaction is, in absolute terms, a reasonably fast reaction with reaction rate constants in the order between 10$^{-1}$--10$^{0}$ s$^{-1}$; i.e. it can proceed in astronomical timescales in the absence of competitive reactions. However, we will show in Section \ref{sec:HNOH}, that this reaction channel is minor compared with the competitive \ce{H2} abstraction reactions due to the increased influence of quantum tunnelling in the latter. 

\begin{figure}
    \centering
    \includegraphics[width=\linewidth]{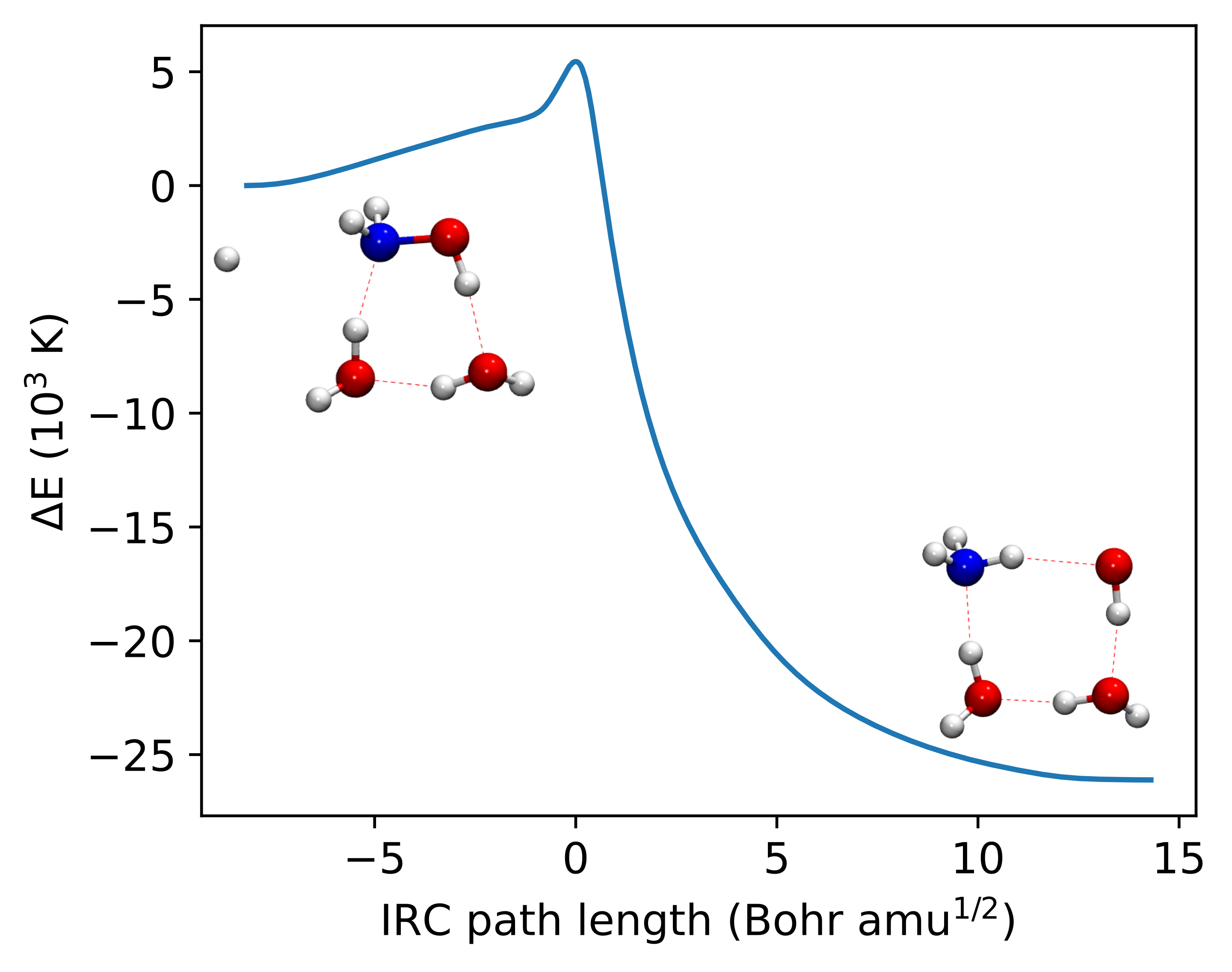}
    \caption{Intrinsic reaction coordinate calculations for reaction \eqref{eq:res1}. The IRC path is neither corrected for ZPVE contributions nor DLPNO-CCSD(T) energies.}
    \label{fig:irc}
\end{figure}

From a chemical point of view, it is important to ascertain that reaction \eqref{eq:res1} proceeds in a single elementary step, meaning that no intermediate reaction that can be long-lived on an ASW surface is formed during the reaction. We confirmed this behaviour using the transition state structure's intrinsic reaction coordinate (IRC) calculations. The results for the IRC are presented in Figure \ref{fig:irc}. Note that the path length to products is notoriously considerable but leads without any additional barrier to the bimolecular products \ce{NH3} and \ce{OH}. 

\subsubsection{\ce{NH2OH + H -> H2O + NH2}} \label{sec:H2O}

\begin{figure}
    \centering
    \includegraphics[width=\linewidth]{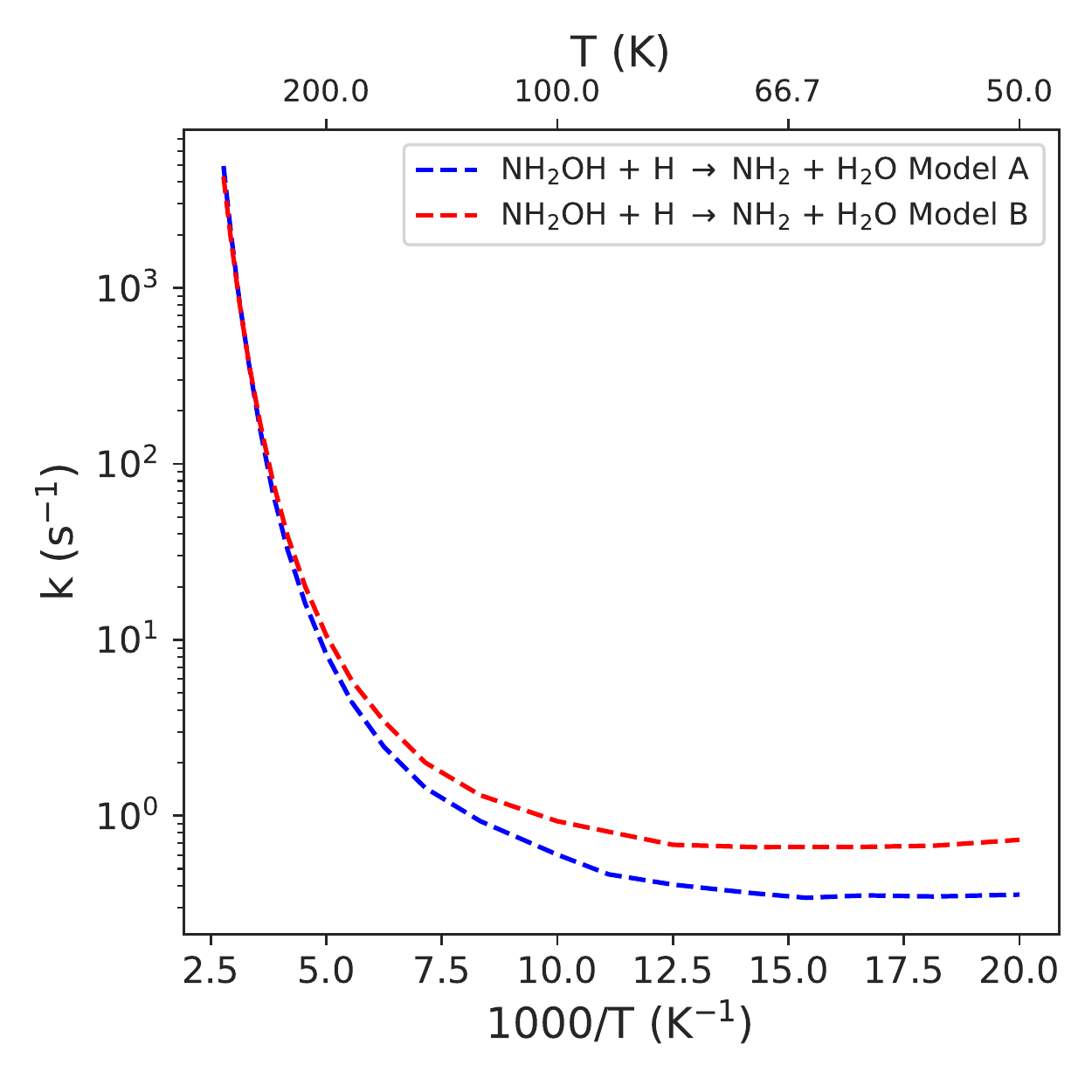}
    \caption{Instanton rate constants for reaction \eqref{eq:res2} in Model \textbf{A} and Model \textbf{B}. The crossover temperature is 406 K for the former and 399 K for the latter.}
    \label{fig:rates_h2o}
\end{figure}

The second reaction we focus on in this work is reaction \ref{eq:res2}. Similar to the case found in Section \ref{sec:NH3} for \ce{NH3}, once hydrogen adds to the OH moiety of \ce{NH2OH}, it breaks it. Subsequent reactions to reform \ce{NH2OH}, through for example the back-reaction:

\begin{equation}
    \ce{H2O + NH2 -> NH2OH + H} \label{eq:h2o_nh2_2}
\end{equation}

\noindent are endothermic (provided that the forward reaction is exothermic), proceed through high activation barriers or are not competitive with \ce{H} abstraction reactions. The reaction energies for reaction \eqref{eq:res2} are $\Delta U_{R, \text{ModelA}}$=-29,762 K and $\Delta U_{R, \text{ModelB}}$=-29,877 K whereas activation energies were found to be $\Delta U_{a, \text{ModelA}}$=6,437 K and $\Delta U_{a, \text{ModelB}}$=6,206 K. In this case, the effect of the inclusion of explicit water molecules in the model does not lead to significant changes in the energetics of reaction. This is due to the transition state for the reaction being located separated from the H-bonding network of the \ce{NH2OH-H2O} system. Nonetheless, the activation energies for the reaction are much higher for reaction \eqref{eq:res2} than for reaction \eqref{eq:res1}. The instanton rate constants for reaction \eqref{eq:res2} are presented in Figure \ref{fig:rates_h2o}. From them, we determine that in the shallow binding sites defined by Model \textbf{A}, reaction \eqref{eq:res2} is much slower than reaction \eqref{eq:res1}. However, it is interesting to observe that in Model \textbf{B}, the increased activation energy of reaction \eqref{eq:res1}, an effect not present in reaction \eqref{eq:res2}, makes reactions \eqref{eq:res1} and \eqref{eq:res2} competitive. As we will show later, this stabilization is not ultimately relevant to the ISM chemistry of \ce{NH2OH}, because H abstraction reactions dominate the chemistry. Finally, we also confirmed that reaction \eqref{eq:res2} proceeds through a single elemental step, like reaction \eqref{eq:res1} using IRC calculations.

\subsubsection{\ce{NH2OH + H -> H2NO + H2}} \label{sec:H2NO}

Reaction \eqref{eq:res3} involves the abstraction of H from an OH moiety, forming \ce{H2} during the process. In the case of formic acid (\ce{HC(O)OH}) \citep{Molpeceres2022}, we observed that abstracting hydrogen from this position is endothermic and proceeds through high activation energy. For \ce{NH2OH} we found that the abstraction in the OH moiety is exothermic but significantly less so than the H addition counterparts (Reactions \ref{eq:res1} and \ref{eq:res2}). The reaction energies are $\Delta U_{R, \text{ModelA}}$=-13,172 K and $\Delta U_{R, \text{ModelB}}$=-13,444 K. In the case of activation energies, we found the highest gap between Model \textbf{A} or Model \textbf{B} for this reaction with $\Delta U_{a, \text{ModelA}}$=4,206 K and $\Delta U_{a, \text{ModelB}}$=7,439 K. This fact is unsurprising considering that the OH moiety of \ce{NH2OH} is interacting directly with a water molecule of the 2\ce{H2O} cluster. Therefore in Model \textbf{B} the hydrogen atom needs to overcome the combined effect of two processes, first the breaking of a hydrogen bond and second the short-range barrier inherent to the reaction.

\begin{figure}
    \centering
    \includegraphics[width=\linewidth]{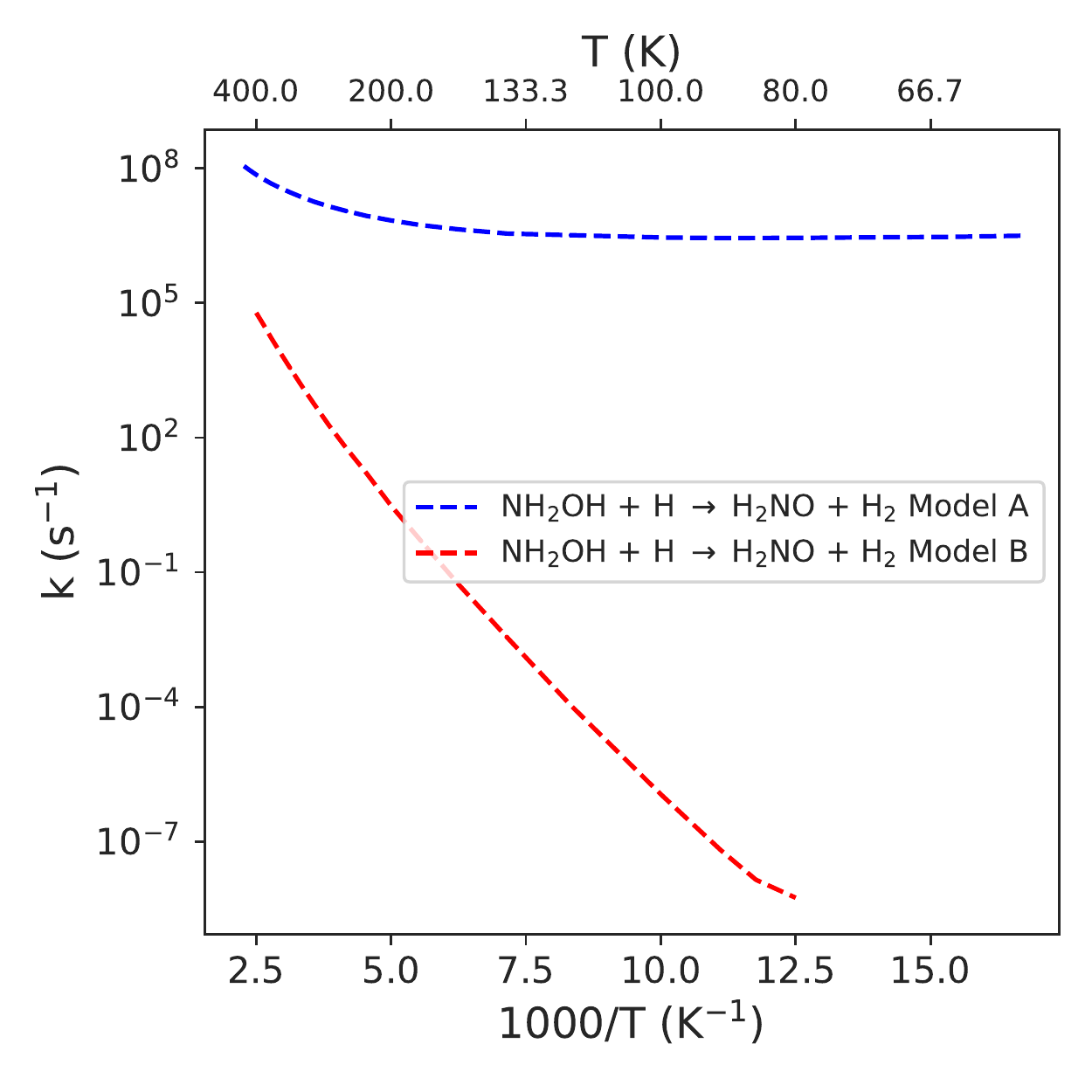}
    \caption{Instanton rate constants for reaction \eqref{eq:res3} in Model \textbf{A} and Model \textbf{B}. The crossover temperature is 539 K for the former and 514 K for the latter.}
    \label{fig:rates_h2no}
\end{figure}

The instanton rate constants for reaction \eqref{eq:res3} are presented in Figure \ref{fig:rates_h2no}. From the plot, we can see the immense effect of the different structural models on the rate constants. While in model \textbf{A} the reaction is the fastest of this study, in model \textbf{B}, we find the slowest reaction. This finding is essentially the same provided in \cite{nguyen_experimental_2019} for the hydrogenation of NO, i.e. a critical relevance of the ice surface in the reaction. We note that for dual-level instantons in Model \textbf{A}, we observed that the rate constant at 50 K is approximately 1.4 higher than at 80 K (not perceptible in the graph). The reason for this unphysical effect can be found in slightly different pre-reactive complex geometries between our energy method  DLPNO-CCSD(T) and our geometry method DSD-PBEP86-D3BJ/def2-TZVP. However, this effect's impact is minute in the overall conclusions.

In section \ref{sec:hno} we discuss the chemical implications of this finding and the current limitations of quantum chemical calculations to cover the diversity of chemically active sites on a real ASW surface. From a purely chemical perspective and for reaction \eqref{eq:res3} we must conclude that the most likely representation of the real situation on a surface must be closer to Model \textbf{B} than to Model \text{A}. Depending on the ice morphology, in the case of, for example, numerous \ce{H2O} dangling bonds on the surface in porous ices \citep{Bossa2015a,mate_spectral_2021}, the situation might be different, and it is hard to predict which model will predominantly describe the reaction.

\subsubsection{\ce{NH2OH + H -> HNOH + H2}} \label{sec:HNOH}

The final reaction with activation energy that we considered is represented in equation \eqref{eq:res4}. This reaction is equivalent to equation \eqref{eq:res3} except that the abstracted H comes from the \ce{NH2} moiety that does not directly interact with the ice because, in interaction with water, the amine group acts as an H-bond acceptor. This effectively leaves the H atoms free to react. The reaction energies are $\Delta U_{R, \text{ModelA}}$=-9,527 K and $\Delta U_{R, \text{ModelB}}$=-10,763 K and the activation energies are $\Delta U_{a, \text{ModelA}}$=3,409 K and $\Delta U_{a, \text{ModelB}}$=3,218 K. It is clear then clear that the major product of reaction of \ce{NH2OH + H} must be \ce{HNOH + H2}, both in Model \textbf{A} and in Model \ce{B}. That is why, despite the reaction energies being lower than in the case of reaction \eqref{eq:res3}, and \ce{H2NO} being the thermodynamical product of the reaction, \ce{H2NO} is not the most likely product of reaction at ISM conditions due to the influence of the water matrix in protecting the -OH moeity from abstraction.

\begin{figure}
    \centering
    \includegraphics[width=\linewidth]{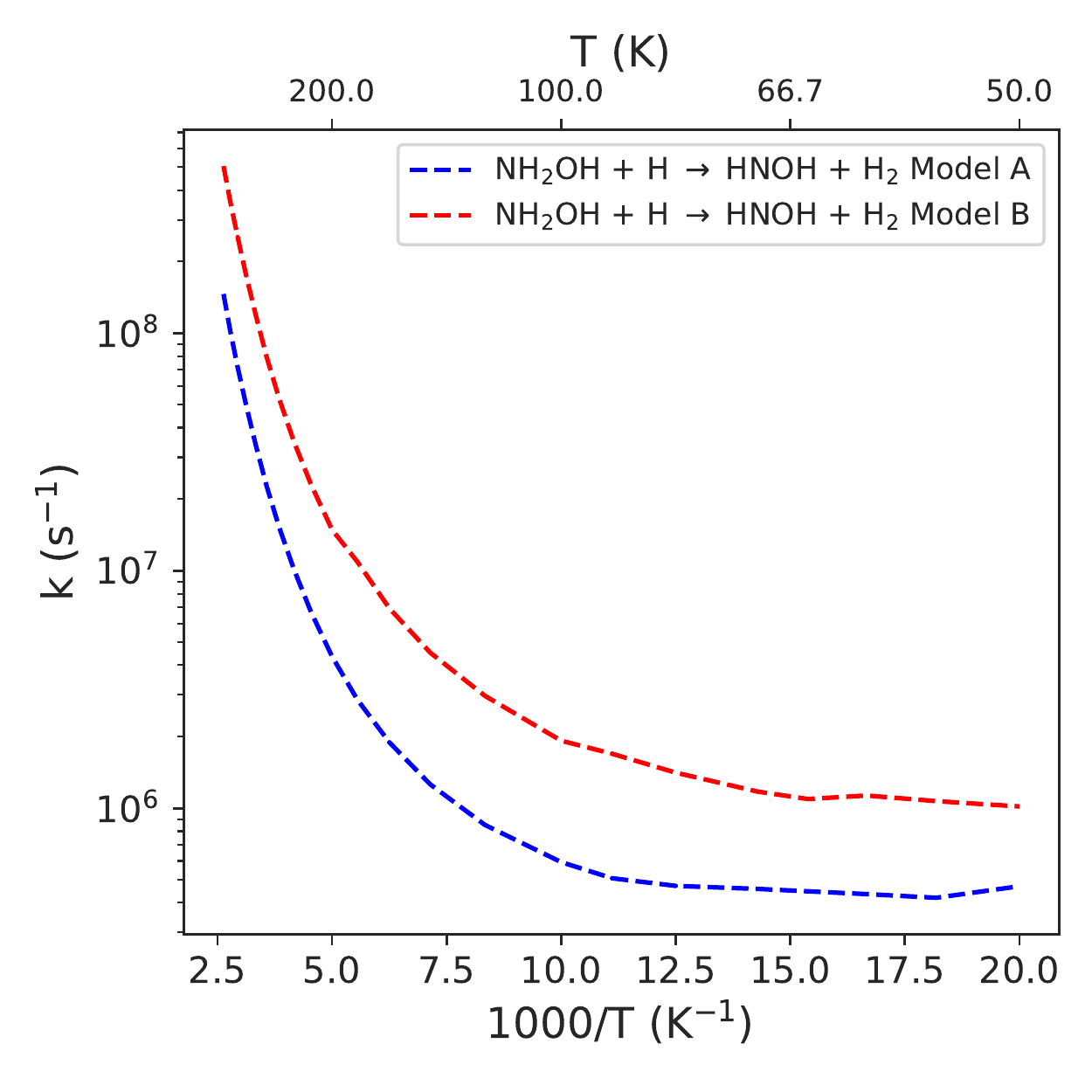}
    \caption{Instanton rate constants for reaction \eqref{eq:res4} in Model \textbf{A} and Model \textbf{B}. The crossover temperature is 420 K for the former and 416 K for the latter. }
    \label{fig:rates_hnoh}
\end{figure}

The instanton rate constants for this reaction are presented in Figure \ref{fig:rates_hnoh} and support the energetic point of view presented above. They are similar both for Model \textbf{A} and Model \textbf{B}. It is important to note that in the gas phase, these rate constants are missing a factor 2 from the rotational symmetry number $\sigma$. In amorphous ices such as the ones these models aim for, such a degeneracy must be broken. Our values for the activation energies of reactions \ref{eq:res2} and \ref{eq:res3} vary with respect to the values in \cite{nguyen_experimental_2019} that reported activation energies of $\sim$ 1,500 K. While such deviations may stem from the computational method, they do not affect the main conclusion of this or their work, that is that \ce{NH2OH} reacts with H \emph{via} abstraction reactions, and mainly from the \ce{NH2} moiety in the presence of strong H-bond networks.

\subsubsection{Hydrogenation of \ce{HNOH} and \ce{H2NO}} \label{sec:brokensym}

\begin{figure}
    \centering
    \includegraphics[width=\linewidth]{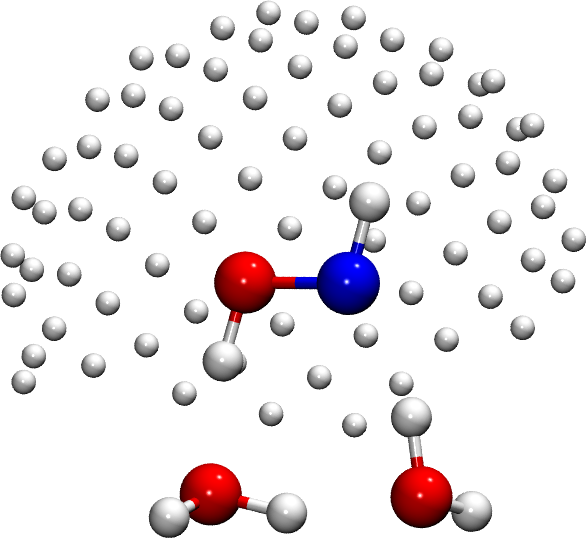}
    \caption{Initial positions for the sampling of reaction outcomes in radical-radical recombinations (Reactions \eqref{eq:brok:1} and \eqref{eq:brok:2}) for Model \textbf{B}. The image shows 77 initial structures of the 150 considered for this reaction. White spheres represent initial poisions (one at a time) of the hydrogen atoms.}
    \label{fig:sampling}
\end{figure}

Reactions \eqref{eq:res1} and \eqref{eq:res2} are irreversible in the context of the destruction of hydroxylamine by H atoms. As we discussed in Section \ref{sec:NH3}, there is not, to the best of our knowledge, a significant production of \ce{NH2OH} from \ce{OH + NH3} or \ce{NH2 + H2O} (Reactions \eqref{eq:h2o_nh2} and \eqref{eq:h2o_nh2_2}). However, this is not the case for \ce{HNOH} or \ce{H2NO}, both radicals susceptible to reacting \emph{via} radical-radical recombination. In the same vein as for the destruction reactions, we have sampled the reactions reforming \ce{NH2OH} from H atoms, reactions \ref{eq:brok:1} and \ref{eq:brok:2}. An example of the sampling for \ce{HNOH}, Model \textbf{B} is depicted in Figure \ref{fig:sampling}.

Involving radical-radical recombinations, for this type of reaction we evaluate the branching ratio of the reaction, which is the defining quantity to input in astrochemical models. As we mentioned in Section \ref{sec:met:react} we sampled 150 initial positions for each chemical model (Model \textbf{A} and \textbf{B}). We must mention that in some of our other works \citep{Molpeceres2021b, Miksch2021} not all the geometry optimizations that we sampled ended in a reactive event due to the flatness of the PES and the location of spurious minima. For all three reactive trajectories and irrespective of the structural model, the reactions that we found are:

\begin{align}
    \ce{HNOH + H &-> NH2OH} \label{eq:brok:result:1} \\ 
    \ce{H2NO + H &-> NH2OH} \label{eq:brok:result:2} \\ 
    \ce{H2NO + H &-> NH3O}. \label{eq:brok:result:3}
\end{align}

For Model \textbf{A} and reactions \eqref{eq:brok:result:2} and \eqref{eq:brok:result:3}, we found roughly a 40/60 \% ratio in favor of reaction \eqref{eq:brok:result:3}. The same reactions in Model \textbf{B} yield 45/55 \%. For HNOH, reaction \ref{eq:brok:result:1} dominates in 100\% of the reaction events, both for Model \textbf{A} and \textbf{B}.

The zwitterion \ce{NH3O}, ammonia oxide, was briefly mentioned in \cite{tsegawFormationHydroxylamineLowTemperature2017}, indicating that is less stable than hydroxylamine, with an energy gap of around 1 eV ($\sim$ 11,600 K, \citealt{wangThermalDecompositionPathways2010}). While the formation of \ce{NH3O} in our simulations resembles a non-equilibrium situation and the actual relative energy between the structural isomers is not crucial in this context, we hypothesize that \ce{NH3O} undergoes very fast isomerization on water ice, following, for example, a proton shuttle mechanism \citep{Molpeceres2021c, Perrero2022}. In the following sections, we assume a rapid \ce{NH3O ->  NH2OH} conversion, and we do not deepen in the interaction of \ce{NH3O} with water ices.

It is important to mention that both in reactions \eqref{eq:brok:result:1} and \eqref{eq:brok:result:2} a mixture of cis-\ce{NH2OH} and trans-\ce{NH2OH} is formed, but in light of the results obtained in Section \ref{sec:isomerism}, we conclude that the isomerization must take place in very short timescales, as shown by the blue solid line (for gas) and blue dashed line (for surface isomerization) in Figure \ref{fig:rates_isomerization}. 

\subsection{\ce{NH2OH}, \ce{HNOH}, and \ce{H2NO} binding energies}

\begin{figure*}
    \centering
    \includegraphics[width=5cm]{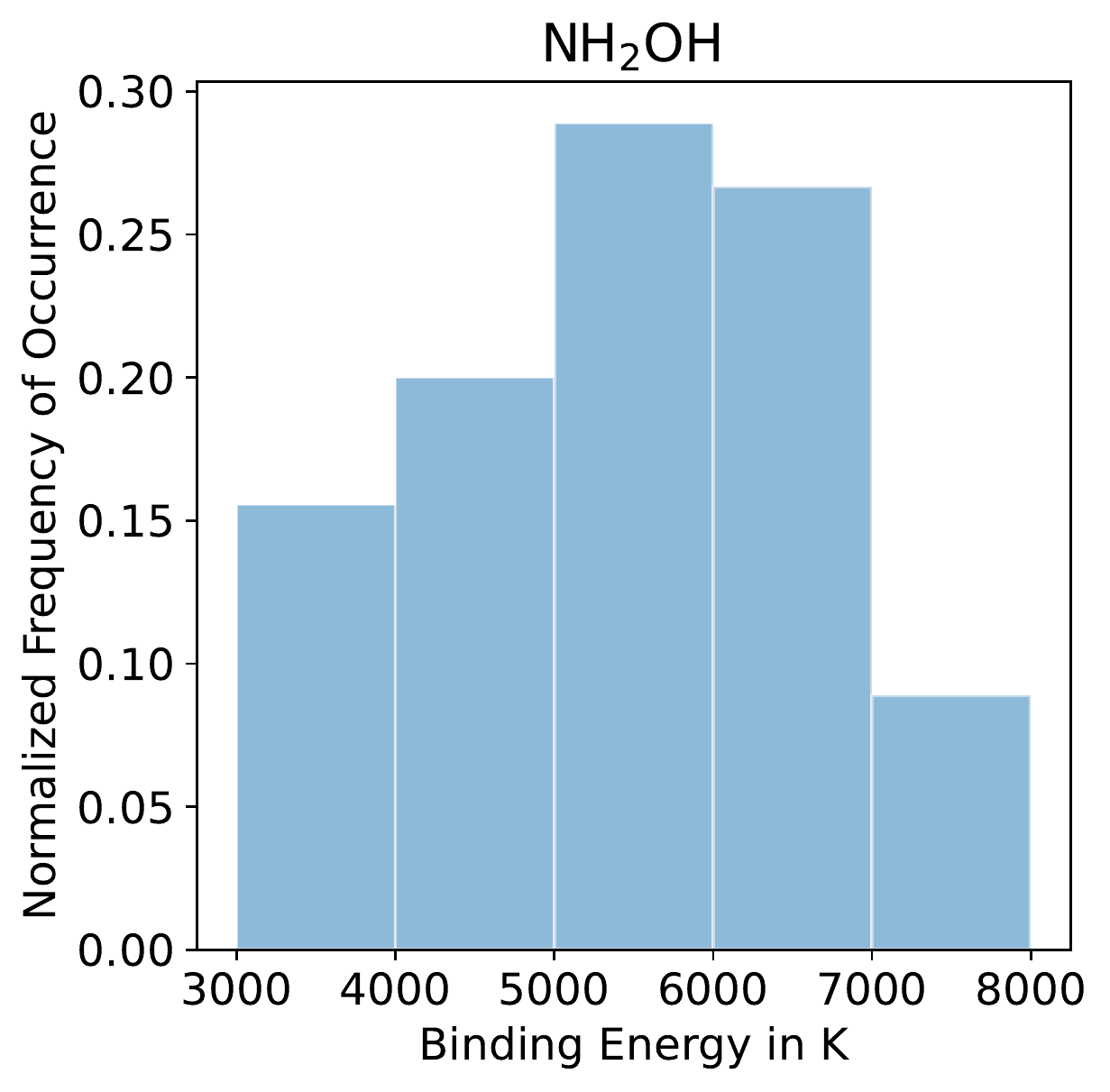}
    \includegraphics[width=5cm]{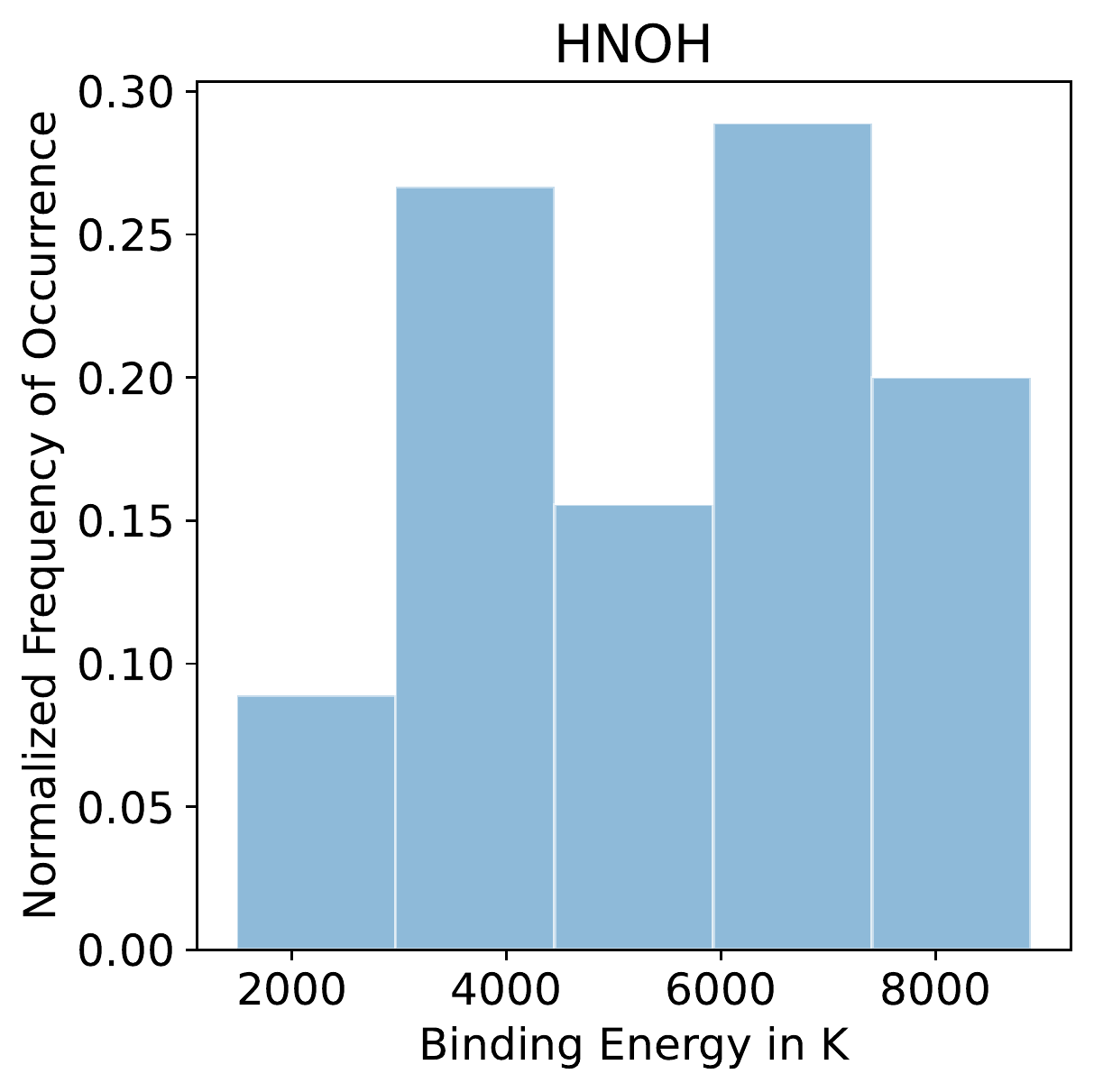}
    \includegraphics[width=5cm]{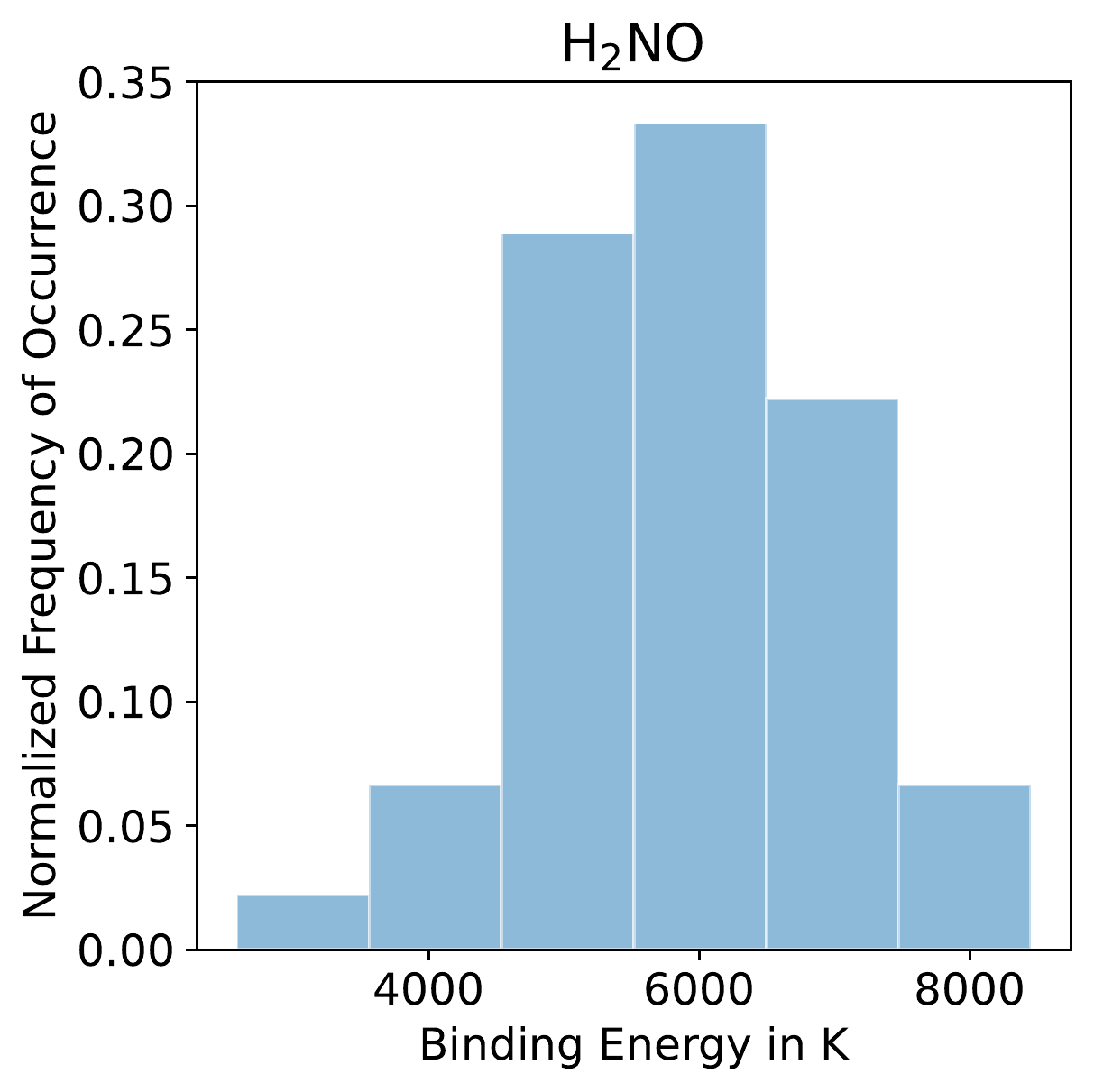}
    \caption{Binding energy distribution for the main intermediates and \ce{NH2OH} of the present study. (Left). Binding energies of \ce{NH2OH}. (Center). Binding energies of \ce{HNOH}. (Right). Binding energies of \ce{H2NO}.  }
    \label{fig:binding}
\end{figure*}

Binding energies for \ce{NH2OH} were calculated to include a refined guess into astrochemical models. The motivation for updating these binding energies in the context of hydroxylamine destruction discussion can be found in \cite{Wakelam2017} (to the best of our knowledge, the only study explicitly addressing \ce{NH2OH}) where the authors state that they found \ce{NH2OH} binding to be likely overestimated due to the computational protocol that they applied (that could not account for the donor-acceptor nature of the hydroxylamine molecule) with reported values in between of 6000 K to 7400 K, depending on the theoretical model. Our results for the average binding energies using high-level quantum chemical calculations are presented in Table \ref{tab:binding}, and the binding energy distributions can be found in Figure \ref{fig:binding}. From both the distributions and the average values, we confirm that the assumption presented in \cite{Wakelam2017} was correct and that there was an overestimation of the \ce{NH2OH} binding energies on water ice. However, we do not expect this deviation to influence their study focused on the role of binding energies on chemical desorption since \ce{NH2OH} remains less volatile than water. 

\begin{table}
    \caption{ Average binding energies (in K) of the main intermediates and \ce{NH2OH} derived in the present study.}
    \label{tab:binding}
    \centering
    \begin{tabular}{cc}
    \hline
    Molecule & Binding Energy (K)  \\
    \hline
    \ce{NH2OH} & 5382 \\
    \ce{HNOH}  & 5536 \\
    \ce{H2NO} & 5925 \\
    
    \hline
    \end{tabular}
\end{table}

\section{Chemical Modeling} \label{sec:discuss}

We included our quantum chemically derived activation energies, binding energies and branching ratios into a three-phase astrochemical model (e.g., gas, ice surface, mantle) presented in \cite{furuya_water_2015} to simulate a molecular cloud with limited external energy input. Within this model, we considered two temperature settings. First, 10 K, where hydrogenation reactions must dominate, and second 20 K, where we determined the importance of alternative \ce{NH2OH} formation routes (e.g. ammonia oxidation). We used the chemical network presented in \cite{Garrod2013}, updated by the reactions presented in Section \ref{sec:results}, and in \cite{Congiu2012,nguyen_experimental_2019} and \cite{He2015b}. Further details of the specific chemical reactions are provided later in the text (Section \ref{sec:chemical_model}). The molecular cloud physical model consists of a pseudo-time-dependent model whose physical conditions are presented in Table \ref{tab:conditions_pseudo} and elemental abundances relative to hydrogen nuclei in Table \ref{tab:initial_abundances}. In our model, grain-surface reactions use the modified rate approach of \cite{garrodNewModifiedrateApproach2008a}, Model A. In evaluating the rate of grain-surface reactions with activation barrier, the diffusion-reaction competition is considered \citep{changGasgrainChemistryCold2007}. As defined above, we assume that the diffusion energy of a given adsorbate is $\sim$ 0.4 its binding energy. We set the exceptions in this factor for \ce{H2} (0.5), \ce{H} (0.5), and \ce{CO} (0.3) based in our recent studies \citep{furuya_quantifying_2022,furuyaDiffusionActivationEnergy2022}. In our model, chemistry on the grain takes place in the outermost four layers of the dust particle, with a chemically inactive mantle. Quantum tunnelling of atomic H is taken into account in the evaluation of the diffusion-reaction mechanism. We explicitly include our own derived binding energies for \ce{NH2OH}, \ce{H2NO} and HNOH into the model. Because of the high binding energy of \ce{NH2OH} and related molecules, non-thermal mechanisms returning the molecule to the gas phase are paramount. In our models, we considered photodesorption, from the primary interstellar UV fields as well as the secondary UV field of molecular clouds \citep{Prasad1983}, with a photodesorption yield of 1$\times$10$^{-3}$ for all molecules. We also considered cosmic-ray-induced desorption due to the stochastic heating of cosmic dust in its Hasegawa-Herbst formulation \cite{Hasegawa1993}, with a peak-temperature per collision event of 70 K for 10 $\mu$s. Finally, we also included chemical desorption in our models. For that purpose, we considered two possible scenarios. First, a model with a constant desorption probability per chemical reaction of $\sim$ 1\% \citep{garrodNonthermalDesorptionInterstellar2007}. In the second model, we neglected chemical desorption. While the latter model is more physically sound for a molecule as strongly bound as \ce{NH2OH}, the former allows for an increased return of the \ce{NH2OH} molecule formed on grains to the gas phase of molecular clouds. It is important to mention that in the galactic centre, where \ce{NH2OH} was recently detected \cite{Rivilla_2020}, an additional mechanism operates, returning molecules to the gas-phase that are mechanical shocks induced by cloud-cloud collisions, the explicit simulation of this mechanisms is currently out of the scope of our models. 

During our chemical simulations, we found that the reaction rate of \ce{HNO + H -> NO + H2} is key for the evolution of \ce{NH2OH}. Thus we ran a grid of models varying its branching ratio concerning the competing reactions. We discuss these findings in \ref{sec:hno}. Likewise, we further analyzed the viability of \ce{NH3 + O -> NH2OH} in Section \ref{sec:20K}. 

\begin{figure}
    \centering
    \includegraphics[width=8cm]{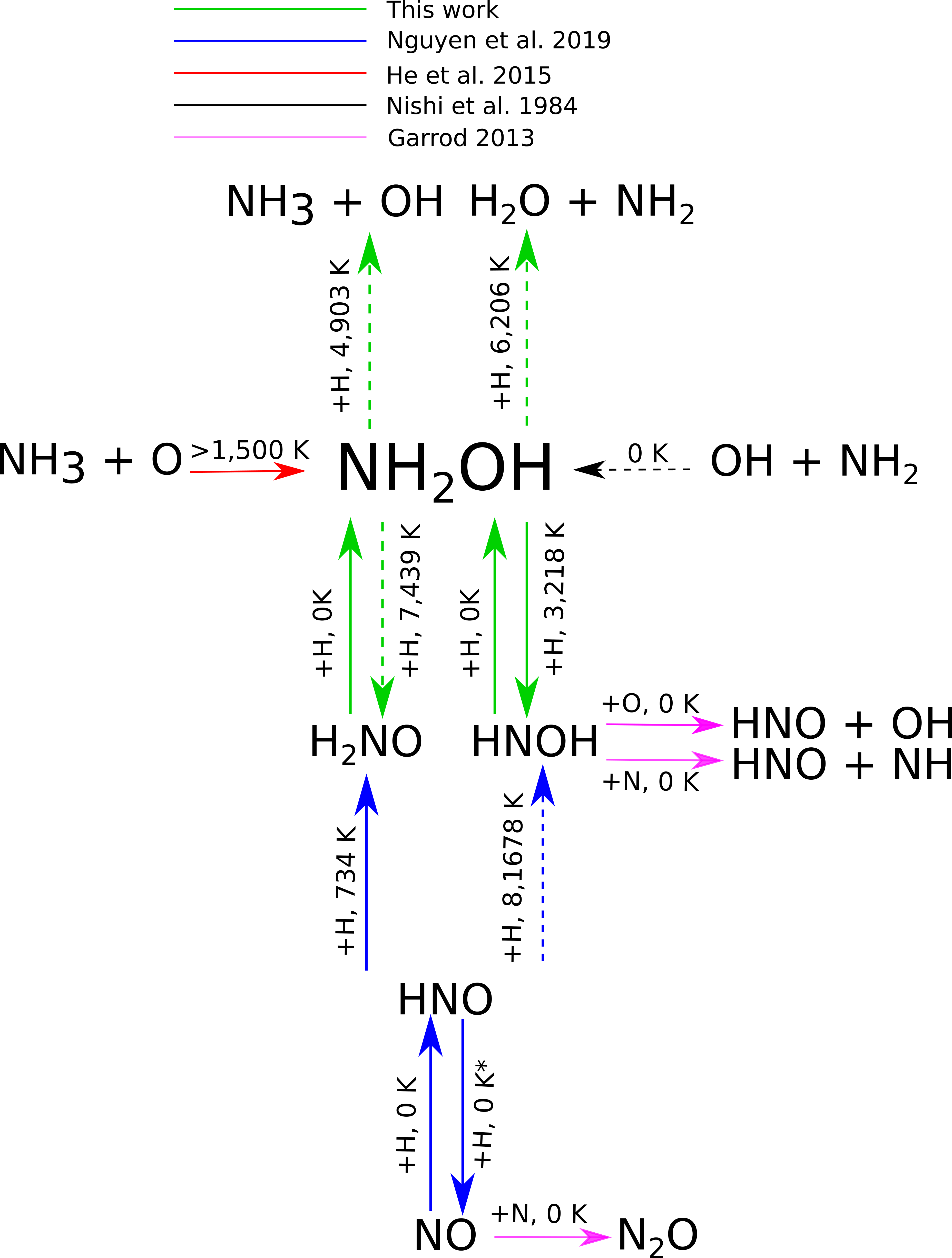} \\

    \caption{Schematic view of the formation and destruction paths of \ce{NH2OH}. Some alternative reactions not leading to hydroxylamine are excluded in the scheme (See text and Table \ref{tab:new_reactions}). In solid line, main reaction routes contributing to the reaction network, either at 10 K (Sections \ref{sec:chemical_model} and \ref{sec:hno}) or at 20 K, section (\ref{sec:20K}). In dashed lines, reactions whose contribution is minor at all temperatures.}
    \label{fig:scheme}
\end{figure}

Finally, we did not check the influence of the chemical structural model (Model \textbf{A} and \textbf{B}) of our quantum chemical calculations in the abundances for the molecular cloud model. We assume Model \textbf{B} to be more realistic and base the astrophysical conclusions on it. Hence, we continue with the values in Model \textbf{B} throughout this section, indicating that a small contribution from binding sites closer to Model \textbf{A} may influence variability in our results. We enumerate the main reactions added to \citet{Garrod2013}'s reaction network in Table \ref{tab:new_reactions} and a scheme of the main reactions is portrayed in Figure \ref{fig:scheme}. 

\begin{table}
    \caption{ Initial physical conditions utilized for modelling the evolution of \ce{NH2OH} under molecular cloud conditions.}
    \label{tab:conditions_pseudo}
    \centering
    \begin{tabular}{cc}
    \hline
    Parameter & Value  \\
    \hline
    n$_{\text{gas}}$  & 2x10$^{4}$ cm$^{-3}$    \\
    A$_{v}$           & 5 mag                   \\
    $\zeta$           & 1.3x10$^{-17}$ s$^{-1}$ \\
    T$_g$             & 10,20 K                    \\
    T$_d$             & 10,20 K                    \\
    
    \hline
    \end{tabular}
\end{table}

\begin{table}
    \caption{ Initial elemental abundances with respect to H nuclei. Taken from \citet{Aikawa1999}.}
    \label{tab:initial_abundances}
    \centering
    \begin{tabular}{cc}
    \hline
    Element  & Initial Abundance  \\
    \hline
    \ce{H2}  & 0.5    \\
    \ce{He}  & 9.8$\times10^{-2}$ \\ 
    \ce{N}   & 2.5$\times10^{-5}$ \\
    \ce{O}   & 1.8$\times10^{-4}$ \\
    \ce{C+}  & 7.9$\times10^{-5}$ \\
    \ce{S+}  & 9.1$\times10^{-8}$ \\
    \ce{Si+} & 9.7$\times10^{-9}$ \\
    \ce{Fe+} & 2.7$\times10^{-9}$ \\
    \ce{Na+} & 2.3$\times10^{-9}$ \\
    \ce{Mg+} & 1.1$\times10^{-8}$ \\
    \ce{Cl+} & 2.2$\times10^{-10}$ \\
    \ce{P+}  & 1.0$\times10^{-9}$ \\
    \hline
    \end{tabular}
\end{table}

\begin{table*}
    \caption{ Surface reactions added to \citet{Garrod2013} reaction network and employed in this work. Note that \ce{H2NO} gas-phase chemistry was also included, and we considered it equivalent (e.g. same reactions and rate constants) to the chemistry of HNOH (assuming different products). $\alpha$, $\beta$, $\gamma$ are the branching ratio of the reaction accounting for all possible reaction channels for a reaction, the activation energy in K, and the barrier width in~{\AA} for the tunnelling rate constants. For a more in-depth description of the formalism, we refer to \citet{ruaud_gas_2016}, Section 2.3. An $\alpha$ value of 0 means deactivating the reaction (see text), and the rest of the parameters are not applicable (n/a in the table). }
    \label{tab:new_reactions}
    \centering
    \begin{tabular}{lcccc}
    \hline
    Reaction            & $\alpha$ & $\beta$ & $\gamma$ & Reference \\
    \hline
    \ce{H + HNO -> NO + H2} & 0.00 -- 0.33$^{i}$  &  0     & 0.00     & \cite{nguyen_experimental_2019} \\
    \ce{H + HNO -> H2NO}    & 0.50 -- 0.33$^{i}$ &  734   & 1.00     & \cite{nguyen_experimental_2019} \\
    \ce{H + HNO -> HNOH}    & 0.50 -- 0.33$^{i}$ &  8,167 & 1.00     & \cite{nguyen_experimental_2019} \\
    \ce{HNO + HNO -> N2O + H2O}  & 1.00 &  0 & 0     & \cite{nguyen_experimental_2019} \\
    \ce{HNO + NO -> N2O + OH}  & 1.00 &  0 & 0      & \cite{nguyen_experimental_2019} \\
    \ce{H + H2NO -> HNO + H2} & 0.00 & n/a & n/a      & This work (Section \ref{sec:brokensym}) \\
    \ce{H + HNOH -> HNO + H2} & 0.00 & n/a & n/a      & This work (Section \ref{sec:brokensym}) \\
    \ce{H + H2NO -> NH2OH}$^{ii}$ & 0.50 & 0 & 0.00  &  \cite{nguyen_experimental_2019} \& This work (Section \ref{sec:brokensym}) \\
    \ce{H + HNOH -> NH2OH} & 0.50 & 0 & 0.00   &  \cite{nguyen_experimental_2019} \& This work (Section \ref{sec:brokensym}) \\
    \ce{O + NH3 -> NH2OH}   & 1.00 &  1,500 & 1.00   & \cite{He2015b} \& This work (Section \ref{sec:20K}) \\
    \ce{NH2 + OH -> NH2OH}  & 1.00 &  0 & 0.00  & \cite{Zheng2010} \& \cite{nishi_photodetachment_1984} \\
    \ce{H + NH2OH -> NH3 + OH}   & 0.25 &  4,903$^{iii}$ & 1.00    & This work (Section \ref{sec:NH3}) \\
    \ce{H + NH2OH -> H2O + NH2}  & 0.25  & 6,206$^{iii}$ & 1.00     & This work (Section \ref{sec:H2O}) \\
    \ce{H + NH2OH -> H2NO + H2}  & 0.25  & 7,439$^{iii}$ &   0.50$^{iv}$  & This work (Section \ref{sec:H2NO}) \\
    \ce{H + NH2OH -> HNOH + H2}  & 0.25  & 3,218$^{iii}$ &   0.50$^{iv}$   & This work (Section \ref{sec:HNOH}) \\
    \hline
    \hline
    \multicolumn{5}{l}{(i) \footnotesize This parameter is varied in some models (See Section \ref{sec:hno})}. \\
    \multicolumn{5}{l}{(ii) \footnotesize \ce{NH3O} is assumed to isomerize to \ce{NH2OH} (See Section \ref{sec:brokensym})} \\
    \multicolumn{5}{l}{(iii) \footnotesize Structural model \textbf{B} values. (See Table \ref{tab:energetics}).} \\
    \multicolumn{5}{l}{(iv) \footnotesize To account for the increased tunneling efficiency of H abstraction reactions.} \\
    \end{tabular}
\end{table*}

\subsection{Molecular cloud model} \label{sec:chemical_model}

\begin{figure}
    \centering
    \includegraphics[width=8cm]{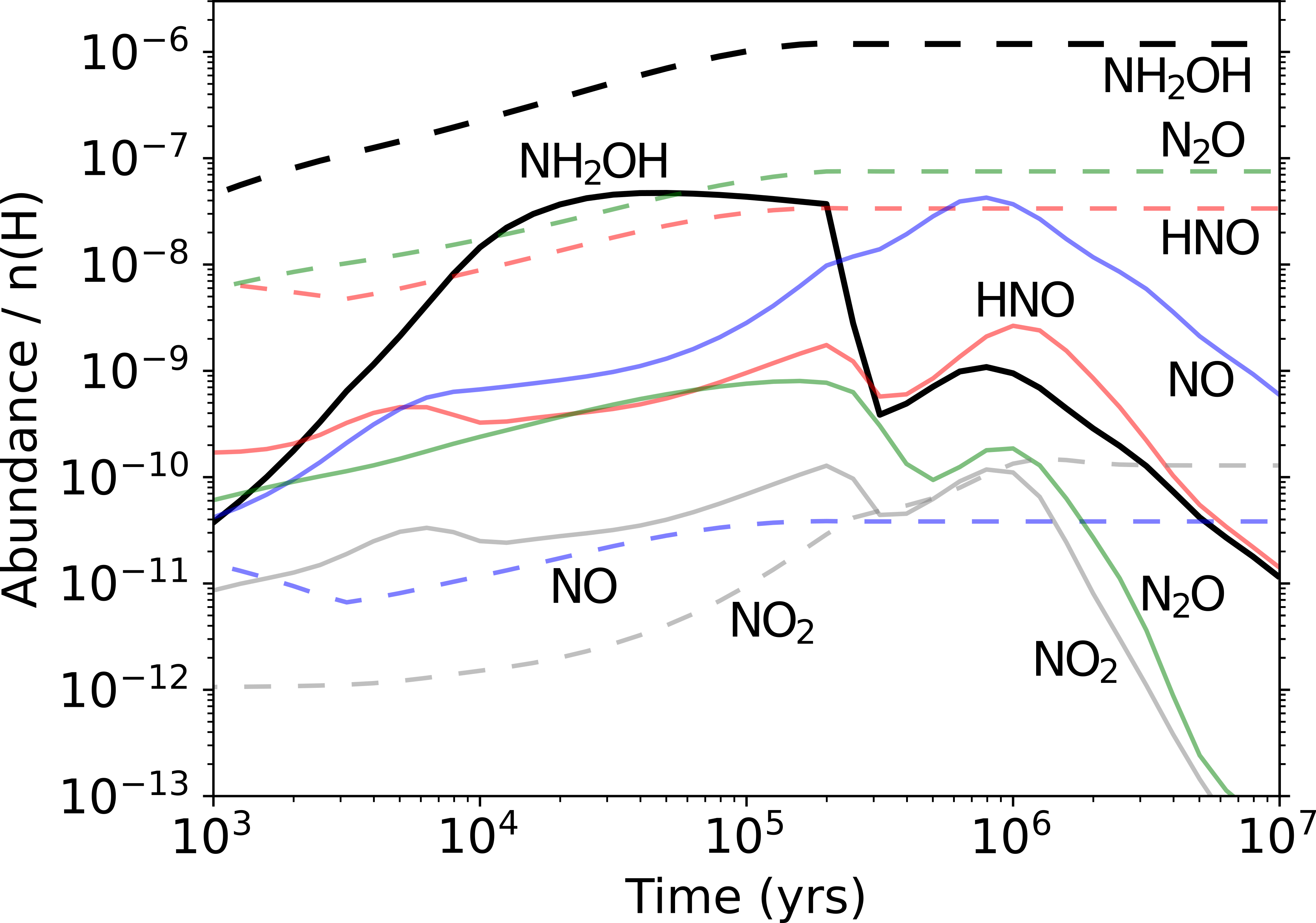} \\
    \vspace{0.5cm}
    \includegraphics[width=8cm]{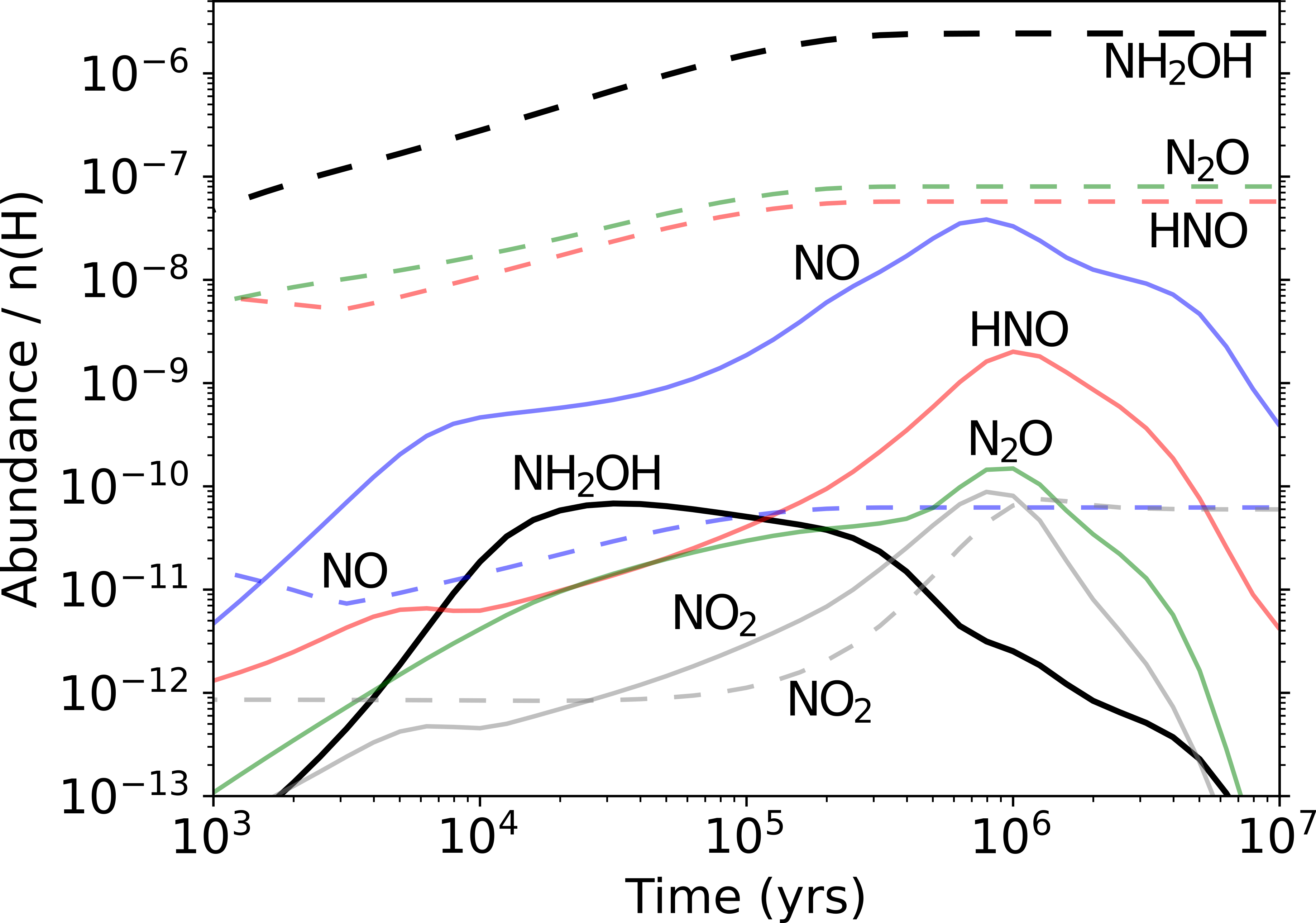} \\
    \caption{(Top). Time-dependent abundances of main NO-bearing molecules for the molecular cloud model at T$_{\text{Dust}}$=10 K and using a constant 0.01 chemical desorption probability. (Bottom). Time-dependent abundances of main NO-bearing molecules for the molecular cloud model at T$_{\text{Dust}}$=10 K and neglecting chemical desorption. Dashed lines represent ice species (surface+mantle), whereas solid lines represent gas phase species. The abundance here is represented to H nuclei (e.g. H + \ce{H2}).}
    \label{fig:molecular:first}
\end{figure}

Figure \ref{fig:molecular:first} shows the abundances of the main NO-bearing compounds for the model, neglecting chemical desorption (bottom panel) and assuming a constant desorption probability (top panel). From the plots, we deduce that for this chemical model, \ce{NH2OH} is the most abundant NO bearing molecule in the ice, formed mainly by reaction \ref{eq:brok:result:2}. This also reflects in a high abundance of \ce{NH2OH} in the gas phase, in the order of a few 10$^{-8}$ at the \ce{NH2OH} peak, which is two orders of magnitude higher than the values reported in \cite{Rivilla_2020}. When we neglect chemical desorption, the abundance in the gas-phase declines to values around 10$^{-11}$, which is lower than the observational value in G+0.693 \citep{Rivilla_2020} by around a factor of two. This observation is unsurprising considering the high binding energy of \ce{NH2OH} that precludes chemical desorption. When chemical desorption is deactivated, most of the \ce{NH2OH} gas abundance is triggered by photodesorption.  When comparing with the upper limit of the abundance presented in \citet{ligterink_alma-pils_2018} of the IRAS16293-2422 hot corino, we observe that the current model predicts the opposite trend with respect the ratio n(\ce{NH2OH})/n(\ce{N2O}) that in our models favours hydroxylamine (gas + ice, since in hot corinos ices are sublimated) in contrast to the observations. The explanation for this discrepancy can be found in the alternative reactions in the hydroxylamine formation reaction network. In particular, the reaction:

\begin{align}
    \ce{HNO + H &-> NO + H2}, \label{eq:NO}
\end{align}

\noindent seems to be of paramount importance and is discussed in the following section.

\subsubsection{The \ce{HNO + H -> NO + H2 reaction}} \label{sec:hno}

Despite the significant uncertainties in the last section's models, they are in line with the values reported in \cite{Garrod2022}, where they report peak abundances in the order of 4$\times$10$^{-6}$. In \cite{Garrod2022}, the authors explicitly mention that the abundance of \ce{NH2OH} may be overestimated. Hence, we think that \ce{NH2OH} ice abundances of 10$^{-6}$, as the ones we obtained in the previous section, are probably unrealistic and a consequence of very efficient hydrogenation on dust grains. To palliate this, we have deepened the study of reaction \ref{eq:NO}. Should this reaction be very efficient then the formation of \ce{NH2OH} would be precluded. This is a general observation of the hydrogenation reaction on ice; more hydrogenations needed to saturate a molecule translate into more possibilities for branches along the hydrogenation sequence.

In their combined theoretical and experimental study \cite{nguyen_experimental_2019} determined that reaction \ref{eq:NO} does not have an intrinsic activation barrier (e.g. it is barrierless), but its occurrence is forbidden due to orientation effects stemming from the HNO-\ce{H2O} interaction. We adhere to this picture, observing the effect of the H-bond network in the activation energies of the reactions explicitly considered in this study (See Table \ref{tab:energetics}). Under ISM conditions, however, when ices are made of more components than \ce{H2O} and/or surface coverage includes more abundant species (e.g. \ce{H2}), it is possible that not all binding modes of \ce{HNO} with the surface preclude reaction \ref{eq:NO}. We emphasize that theoretically, determining the reaction outcomes of interstellar reactions as a function of the appropriate surface composition and surface coverage is unfeasible. Therefore the results in \cite{nguyen_experimental_2019} remain accurate for \ce{H2O} ices. This is the same as saying that in the absence of short-range barriers arising from electronic structure changes in the reactants, e.g. for barrierless reactions, the influence of the surface in the total reaction rate of a species is very hard to gauge from experiments/theory.

A possible way of weighting the importance of a barrierless surface reaction with a marked influence of surface binding, like reaction \ref{eq:NO}, is to run a sensitivity analysis of chemical models to that particular reaction. Since reaction \ref{eq:NO} does not have an entrance barrier, the effect of the surface enters linearly in the rate constant, and it is specifically reflected by the branching ratio of reaction ($\alpha$ in table \ref{tab:new_reactions}). Hence, we determined the importance of reaction \ref{eq:NO} in the production of \ce{NH2OH} by running a grid of models varying the branching ratio of the reaction and accordingly adopting scaled values for the competing reactions:

\begin{align}
    \ce{HNO + H &-> HNOH} \label{eq:HNO1}\\
    \ce{HNO + H &-> H2NO} \label{eq:HNO2}
\end{align}

\begin{figure}
    \centering
    \includegraphics[width=8cm]{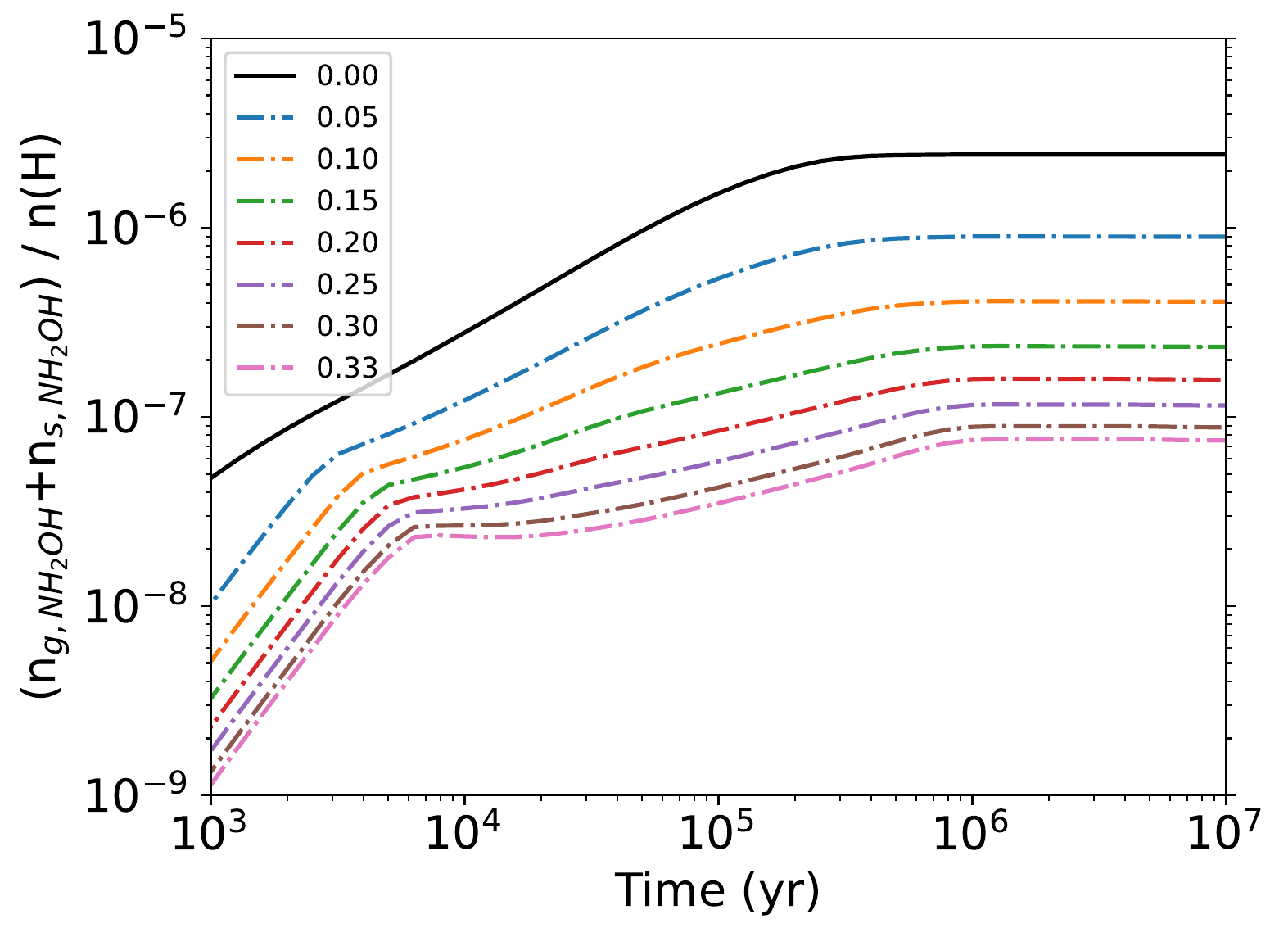}
    \caption{Total abundances of \ce{NH2OH} (e.g. gas + ice) as a function of the branching ratio of reaction for the reaction \ce{HNO + H -> NO + H2}, reaction \ref{eq:NO}. Note that a branching ratio of 0.00 means deactivating the reaction.}
    \label{fig:molecular:sensitivity}
\end{figure}

The sensitivity analysis results are presented in Figure \ref{fig:molecular:sensitivity}. We considered the non-chemical desorption model for this analysis. From the figure, we found that the production of \ce{NH2OH} markedly depends on the efficiency of reaction \ref{eq:NO}, with abundances that span almost two orders of magnitude. The main reason behind the changes in abundances of hydroxylamine lies in the competition between reaction \ref{eq:NO} and reactions \ref{eq:HNO1} and \ref{eq:HNO2}. While reaction \ref{eq:NO} is barrierless and has a low branching ratio, reactions \ref{eq:HNO1} and \ref{eq:HNO2} present an activation barrier, although for reaction \ref{eq:HNO2} this is very low. This establishes a pseudo-equilibrium favoured by the rapid diffusion of H atoms on the surface in the tunneling-mediated reaction-diffusion competition between these reaction channels.

Therefore, we determined that the reaction \ref{eq:NO} is essential and serves as a way of reducing the known problem of the overabundance of hydroxylamine in astrochemical models \citep{Garrod2013, He2015b, Garrod2022}. However, we cannot pinpoint the optimal value of the branching ratio. Should \ce{NH2OH} be detected in cold cores, we will have solid evidence to compare with. Without that, comparisons with the only source where this molecule was detected, G+0.693 \citep{Rivilla_2020}, require careful treatment of the non-thermal mechanisms operating in those environments, which is out of the scope of this work. 

\begin{figure}
    \centering
    \includegraphics[width=8cm]{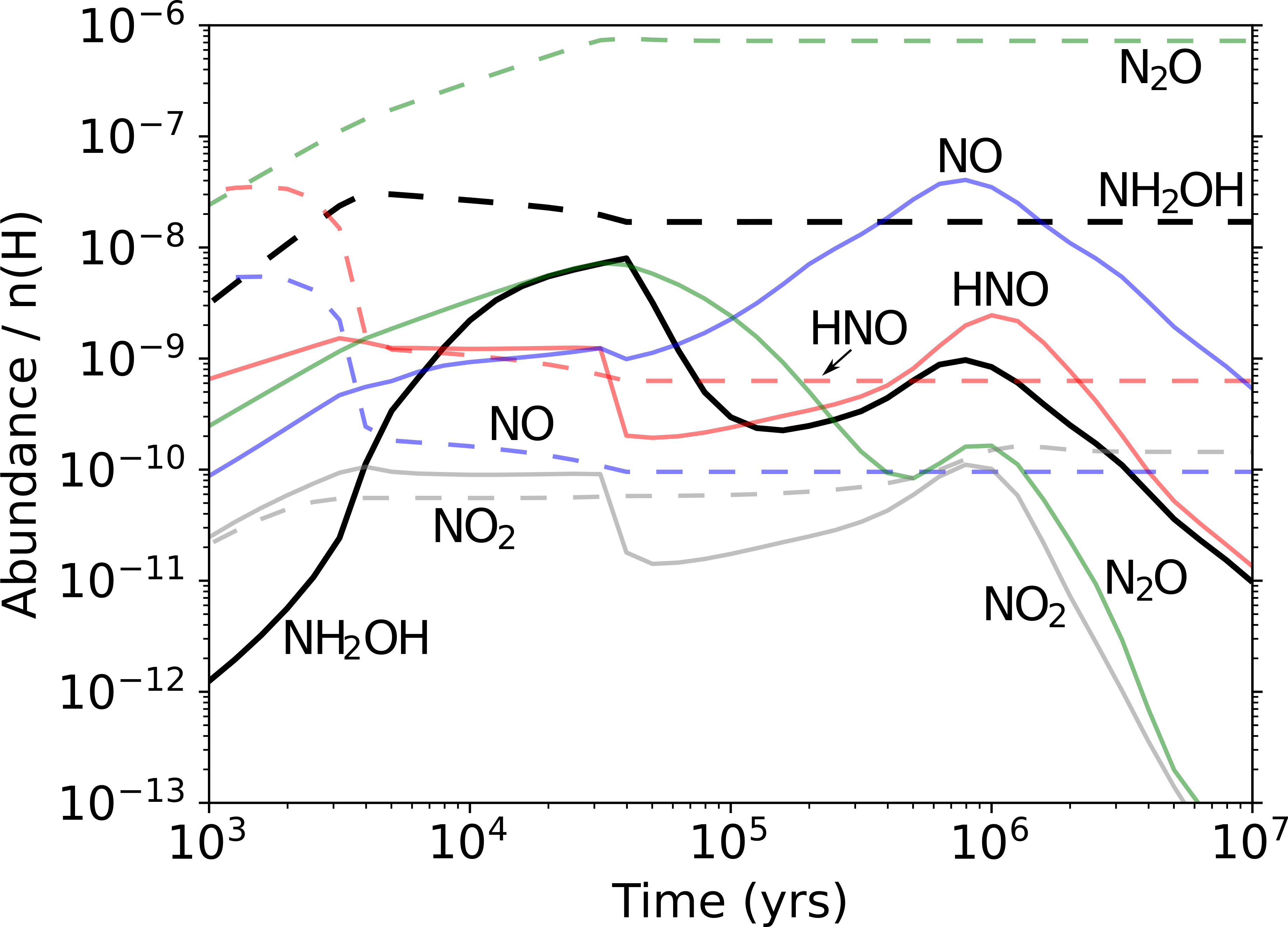} \\
    \vspace{0.5cm}
    \includegraphics[width=8cm]{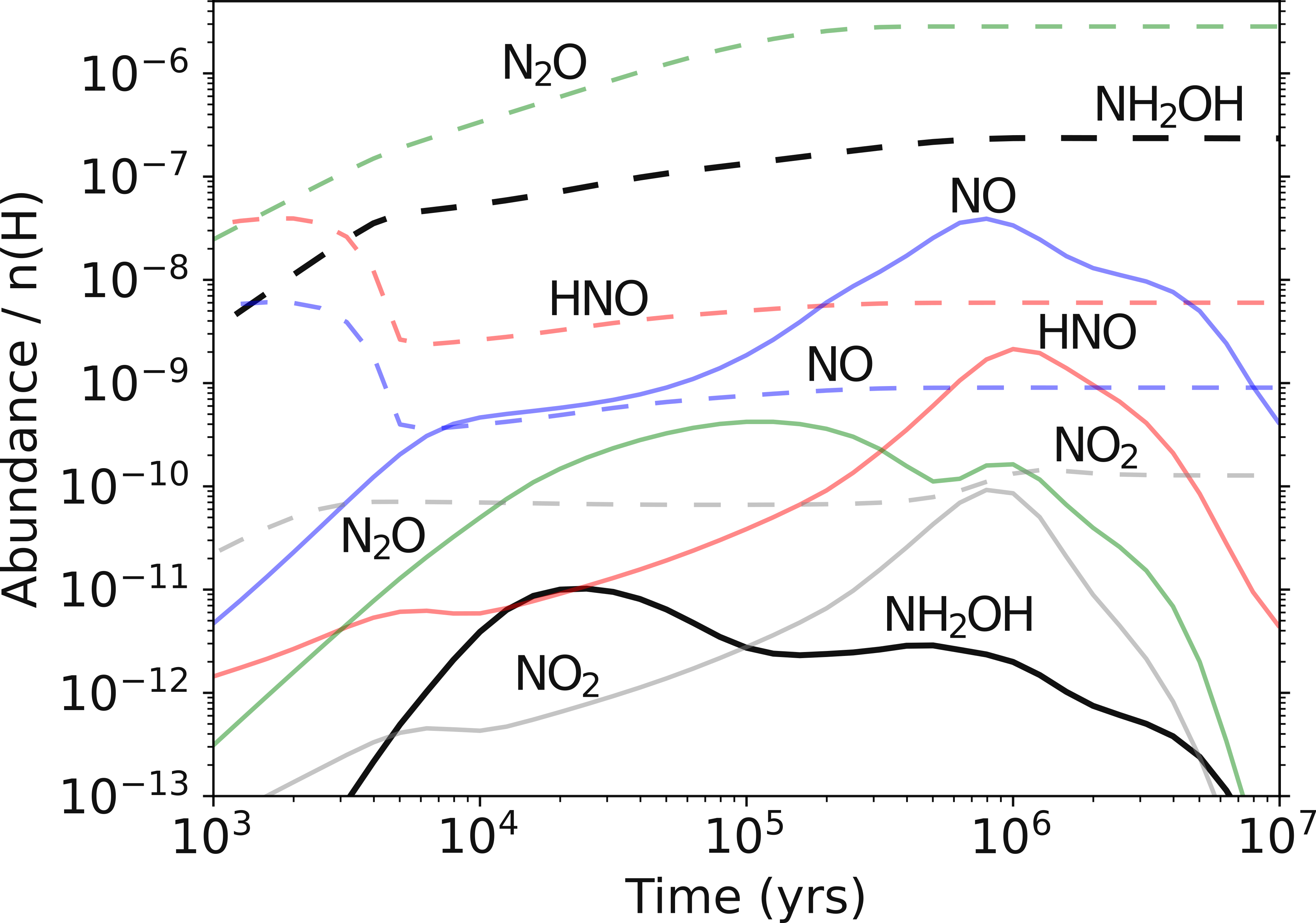} \\
    \caption{Same as Figure \ref{fig:molecular:first} but with branching ratio of 0.15 for reaction \ce{HNO + H -> NO + H2.} }
    \label{fig:molecular:0.15}
\end{figure}

Assuming a compromise value between 0--0.33, for example, 0.15, we have computed the molecular abundances of the important NO-bearing molecules with the constant chemical desorption scheme and neglecting chemical desorption. The results can be viewed in Figure \ref{fig:molecular:0.15}. The change in the branching ratio of reaction \ref{eq:NO} has an important effect on the abundances of both ice and gas \ce{NH2OH}. For the gas, it reduces the peak abundance by around one order of magnitude. More importantly, for the ice's \ce{NH2OH} abundances, we find a decrease of around one or two orders of magnitude in abundance. Such a change promotes the formation of \ce{N2O} through the surface reaction:

\begin{align}
    \ce{NO + N &-> N2O},
\end{align}

\noindent making \ce{N2O} the most abundant NO-bearing molecule on ice and reconciling our models with the lower limit obtained in the IRAS16293-2422 hot-corino \citep{ligterink_alma-pils_2018} in respect to the ratio n(\ce{NH2OH})/n(\ce{N2O}), with the value of the upper limit derived here, that now is correctly accounted for. Concerning G+0.693, our models reproduce well the inferred ratios for nitrogen oxides \citep{Rivilla_2020}, n(NO) \textgreater n(\ce{N2O}) $\sim$ n(\ce{NO2}) \textgreater n(HNO) $\sim$ n(\ce{NH2OH}), albeit with some discrepancies. This trend is reasonably reproduced in our models, which also predict a prevalence of NO as the central nitric oxide at the NO molecules peak (10$^{6}$ yr) and an on-par abundance of \ce{N2O} and \ce{NO2}. However, our model overproduces gas-phase HNO in comparison with G+0.693. Although a direct comparison with G+0.693 is out of the scope of this work, it is worth mentioning that, in the absence of chemical desorption (without the significant contribution of grain chemistry, where the molecule is formed through \ce{NO + H -> HNO}), gas phase HNO is produced in our model from \ce{NH2 + O -> HNO + H} and \ce{H2NO+ + e- -> HNO + H}. New reaction routes may help reduce this molecule's abundance in the gas phase, bringing together models and observations for this particular molecule in the gas of this source. Concerning the absolute abundance of the parent molecule NO, we observe that the peak abundance of gas NO is n(NO)/n(\ce{H2})$\sim$1$\times$10$^{-7}$, in models with and without chemical desorption, and compare relatively well with the abundances of G+0.693 in \cite{Zeng2018} of n(NO)/n(\ce{H2})$\sim$3$\times$10$^{-7}$. We note that this value is rather consistent with the abundances of the Sgr B2(N) molecular cloud as well \citep{Zeng2018}, and previously in other molecular clouds such as L134N, where NO abundances were found to be in between 6$\times$10$^{-8}$ \cite{McGonagle1990} and 3$\times$10$^{-7}$\citep{Gerin1992}. We reiterate that a value of 0.15 for the branching ratio is taken here arbitrarily and is presented for illustrative purposes to highlight the importance of activating or deactivating reaction \ref{eq:NO} and provide a solution to the overabundance of \ce{NH2OH} on ices \citep{Garrod2013, He2015b, Garrod2022}. For this reason, we keep this choice in the models discussed in Section \ref{sec:20K}. 

We found a general explanation for the reduction of \ce{NH2OH} at low temperatures in the combination of three different surface reactions. The most critical reaction is H abstraction from HNO (Reaction \ref{eq:NO}), as explained above, but the H abstraction (Reaction \ref{eq:res4}), and hydrogen addition (Reaction \ref{eq:brok:1}) loop is also important. At low temperatures, the H abstraction reactions dominate the processing of surface \ce{NH2OH}, with reactions rates up to six orders of magnitude higher than other mechanisms such as photodissociation, either by primary or secondary UV, extracted from rates at 1 Myr. To put it in context, photodissociation and processing through reaction \ref{eq:res1} have similar reaction rates. Finally, it is important to consider the non-thermal effects that can return the molecules to the gas. Especially useful for it is a careful observation of the reaction rates for the model including chemical desorption. The comparison of the reaction rates at 1 Myr, between photodesorption (A$_{\text{v}}$=5 mag) and chemical desorption for this model, reveals that chemical desorption dominates the release of molecules to the gas phase. 

The difference between reactions \ref{eq:res3} and \ref{eq:res4} confirms yet another important finding. The HNOH radical is produced by Reaction \ref{eq:res4}, because the activation barrier from \ce{HNO} (Reaction \ref{eq:HNO1}) is very high \citep{Nguyen2020}. This is indirect evidence that the H-abstraction and H-addition loop enhances the chemical desorption rates of \ce{NH2OH} by nearly two orders of magnitude, a value obtained from the comparison between the rates for the \ce{NH2OH} chemical desorption coming from the HNOH radical (only produced by Reaction \ref{eq:res4}) and the chemical desorption from the \ce{H2NO} radical (produced mainly through HNO hydrogenation, reaction \ref{eq:HNO2}). Chemical desorption is likely the main desorption mechanism in cold environments for \ce{NH2OH} and other molecules that can experience H-abstraction/H-addition loops even if the chemical desorption per reactive event is lower than 1\%. Other examples of such loops can be found in \ce{H2S} \citep{Oba2018,Furuya2022} or \ce{PH3} \citep{Nguyen2020,Nguyen2021}, both molecules with positive confirmation of chemical desorption from ASW.

\subsubsection{Model for T$_{\text{Dust}}$=20 K} \label{sec:20K}

We concluded our study of the molecular cloud model by evaluating the influence of alternative reactions to \ce{NO} hydrogenation in the formation of \ce{NH2OH}. As pointed out in \cite{He2015b}, a plausible formation route for \ce{NH2OH} formation is the oxidation of ammonia \emph{via}:

\begin{equation}
    \ce{NH3 + O -> NH2OH}, \label{eq:ammonia_oxydation}
\end{equation}

a reaction that does not proceed at 10 K due to the low mobility of the oxygen atom on ASW but that starts to contribute at 12 K onwards, reaching asymptotic abundances with respect to the dust temperature at 20 K, irrespective of the not-known activation energy for the reaction (see Figure 9 of \citealt{He2015b}). We ran our chemical model, with and without chemical desorption included, for a temperature of 20~K. The selection of the activation energy for the reaction is not trivial, and we tried to constrain such values \textit{ab-initio} through two possible mechanisms:

\begin{align}
    \ce{NH3 + ^{3}O &-> ^{3}NH3O -> ^{3}NH2OH -> ^{1}NH2OH} \\
    \ce{NH3 + ^{3}O &-> NH2 + OH -> NH2OH}.
\end{align}

\begin{figure}
    \centering
    \includegraphics[width=8cm]{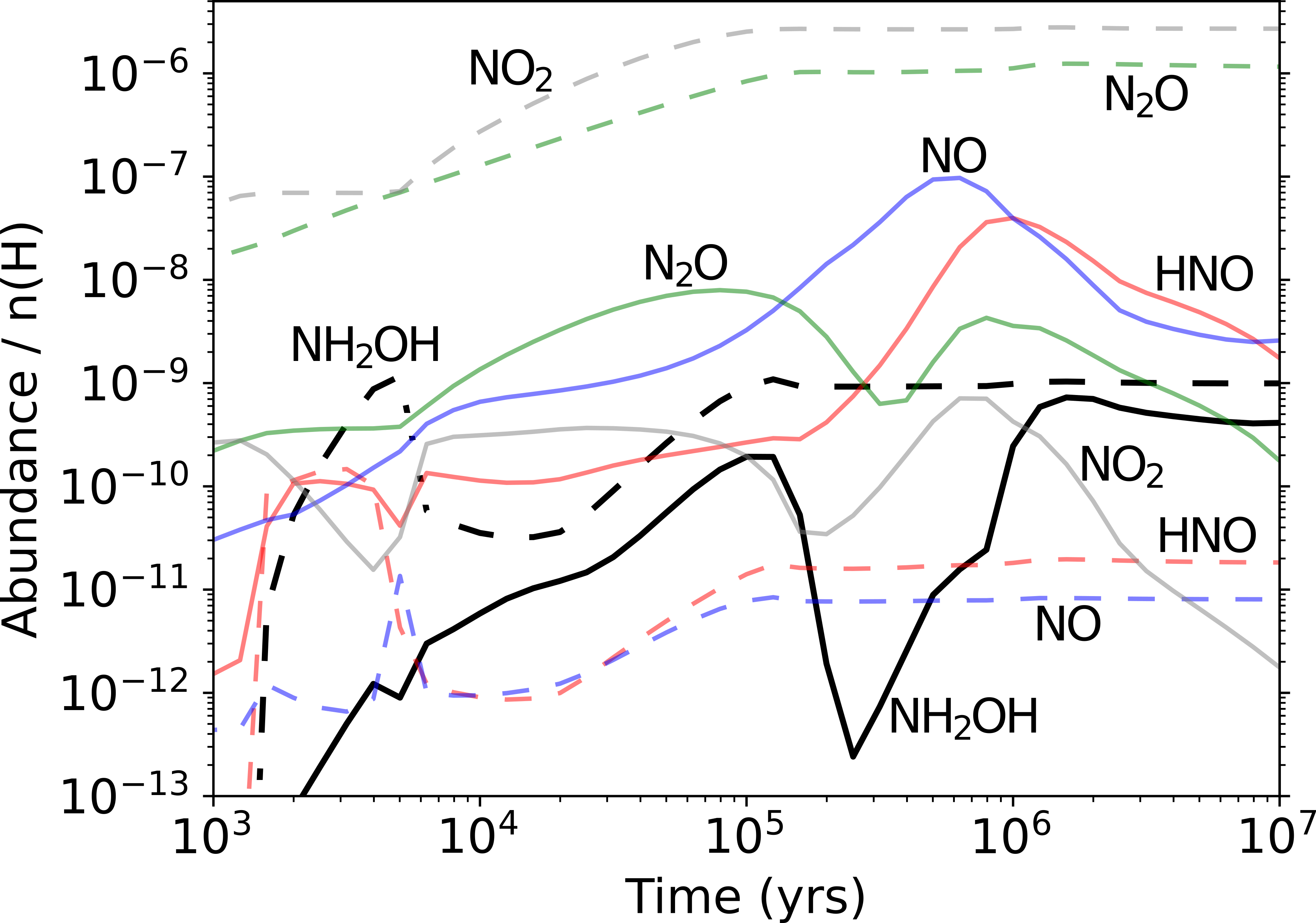} \\
    \vspace{0.5cm}
    \includegraphics[width=8cm]{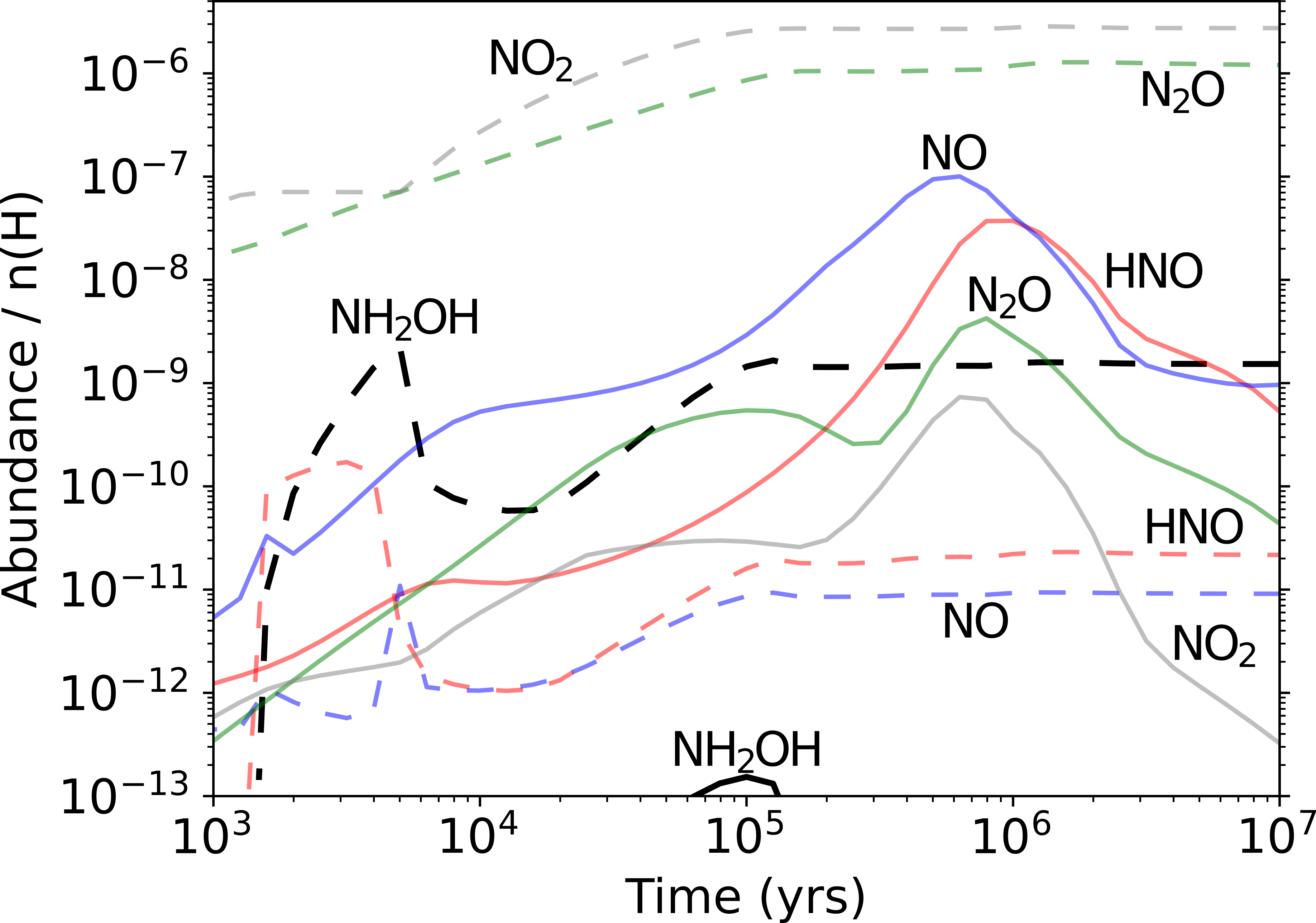} \\
    \caption{Same as Figure \ref{fig:molecular:first} but at T$_{\text{Dust}}$=20 K and branching ratio of 0.15 for reaction \ce{HNO + H -> NO + H2}. }
    \label{fig:molecular:0.15_20K}
\end{figure}

\noindent In the above reactions, and based on our calculations (Model \textbf{B}, DSD-PBEP86-D3BJ/def2-
TZVP. values), \ce{^{3}NH2OH} is a dissociative complex, and the H abstraction in the triplet channel is endothermic. Therefore to explain experimental findings of \cite{He2015b}, an intersystem crossing from the triplet potential energy surface to the singlet one is deemed necessary, and dedicated experimental and quantum chemical studies can be pivotal in understanding the mechanism of reaction \ref{eq:ammonia_oxydation}. In the absence of contamination with $^{1}$O, that was discarded in the experiments above based in prior studies by the same group, \citealt{jingSputteringEffectsWater2013}, such a step must take place in the \ce{^{3}O-NH3} pre-reactant complex. Explicitly simulating that chemical process is out of the scope of this work, but based on this evidence, we think that the (effective) activation energy of reaction \ref{eq:ammonia_oxydation} must be closer to the upper end of the range of values presented in \citet{He2015b}. Therefore, we took 1,500 K for this activation barrier. The results of this model can be found in Figure \ref{fig:molecular:0.15_20K}. Compared to the kinetic models presented in \citet{He2015b}, our final abundances for \ce{NH2OH} are lower by 2-3 orders of magnitude, and the dominant NO-bearing molecules on the surface are \ce{NO2} and \ce{N2O}. This is due to the increased efficiency at 20 K of both H thermal desorption and the barrierless reactions: 

\begin{align}
    \ce{HNOH + O &-> HNO + OH} \label{eq:highT1} \\
    \ce{HNOH + N &-> HNO + NH} \label{eq:highT2}, 
\end{align}

\noindent that are not competitive at 10 K due to the reduced mobility of the N and O atoms to H .The diffusion activation energies of N and O are 288 and 528 K, \emph{e.g} fast diffusion. Therefore, the H abstraction  reactions from \ce{NH2OH} studied in the present work are \emph{also} important at 20 K. They are the onset for destruction routes not operative at lower temperatures. Since reactions \ref{eq:brok:1} and \ref{eq:brok:2} result in H addition rather than abstraction, \ce{NH2OH} is protected thanks to a loop of addition-abstraction reactions that is broken as soon as heavier radicals can diffuse on the dust surface. It would remain to know if reactions \ref{eq:highT1} and \ref{eq:highT2} are barrierless; that is an assumption of our model. Nonetheless, alternative reactions with O or N atoms, not explored in this work, would lead to products different than \ce{NH2OH}, so our conclusion remains valid. Similarly, we expect similar behaviour at higher dust temperatures, providing a sufficiently slow warm-up ramp that allows radical chemistry to proceed. In regards to different molecular ratios, we find that at 20 K, our predicted n(\ce{NH2OH})/n(\ce{N2O}) is significantly different than at 10 K, as well as other ratios such as n(\ce{NH2OH})/n(\ce{NO2}). The variability in these molecular ratios may be helpful in future surveys, interpreting source-to-source variations of \ce{NH2OH}.

\section{Summary and Conclusions} \label{sec:conclusion}

The picture left concerning the prevalence of \ce{NH2OH} by our quantum chemical and kinetic simulations is as follows. As found in \cite{Congiu2012, fedoseev2012, nguyen_experimental_2019}, hydrogenation of NO molecules on dust grains is the primary formation path of \ce{NH2OH} under astrophysical conditions. The reaction is very effective because it is irreversible at low temperatures once the reaction \ce{HNO + H -> H2NO} takes place. However, completely neglecting the another branch \ce{HNO + H -> NO + H2} results in the known problem of the overproduction of \ce{NH2OH} \citep{Garrod2013, He2015b, Garrod2022}. We found that enabling the reaction, even in small amounts (15\% of the available \ce{HNO} binding sites), modulates the reaction enough to have abundances of \ce{NH2OH} close to the values reported in \cite{Rivilla_2020}. However, such an agreement can only be reached when significant influence of non-thermal desorption events is considered, which is not expected under molecular cloud conditions. The  molecular cloud in the galactic centre, G+0.693, the object where this molecule was detected \citep{Rivilla_2020}, is subjected to shocks and increased cosmic-ray fluence that violently releases part of the molecular inventory in grains to the gas. When the conditions of the object are milder, we found that hydroxylamine starts to be processed by heavier radicals and atoms, significantly reducing its abundance. 

Our investigation can benefit from a more in-depth analysis of chemical desorption in the case of hydroxylamine since, in our formulation, we are either neglecting it or, most likely, overestimating it. A rigorous estimation of the likelihood of reactive desorption in the hydroxylamine-water system at 10 K remains to rule out the possibility of finding \ce{NH2OH} in molecular clouds that we consider unlikely in light of our results and at this stage of research. Likewise, warm routes for the surface formation of \ce{NH2OH}, other than the ones considered here, can increase its rate of production, for example, the here unaccounted \ce{NH3 + O2 + CR -> NH2OH + O*} energetic formation route \citep{tsegawFormationHydroxylamineLowTemperature2017}, most likely relevant in ice mantles with enough \ce{O2} abundance. In such a route, \ce{NH2OH} could also be released to gas upon formation, depending on the internal energy of the cosmic-ray dissociated O atoms within the ice. Similarly, the destruction routes of hydroxylamine caused by cosmic-ray analogues can also modulate the \ce{NH2OH} abundance in the grain. We are working on the latter hypothesis to shed light on this particular issue.

With all the findings and caveats presented here, we have given a comprehensive view of the chemistry of a very important prebiotic precursor, and we rationalized some issues concerning its detectability. Further work is needed to unravel the role of this molecule in the gas phase of the ISM, following previous theoretical and experimental results \citep{Largo2004, Barrientos2012, snowGasPhaseIonicSyntheses2007} indicating that it may be a crucial ingredient for the emergence of amino acids and RNA nucleotides in space.

\section*{Acknowledgements}

GM thanks the Japan Society for the Promotion of Science (Grant P22013) and the Alexander von Humboldt Foundation for their support. Some calculations of this work were performed with the support of the state of Baden-Württemberg through bwHPC
and the German Research Foundation (DFG) through grant no INST 40/575-1 FUGG (JUSTUS 2 cluster).
 V.M.R. has received support from the project RYC2020-029387-I (COOL: Cosmic Origins Of Life) funded by MCIN/AEI /10.13039/501100011033. BM thanks Spain MIC grant PID2020-113084GB-I00. KF acknowledges support from JSPS KAKENHI grant Numbers 20H05847 and 21K13967. Y.A. acknowledges support by Grant-in-Aid for Scientific Research (S) 18H05222, and Grant-in-Aid for Transformative Research Areas (A) grant Nos. 20H05847.

\section*{Data Availability}

Data supporting this work will be provided upon request to the corresponding author. 



\bibliographystyle{mnras}
\bibliography{example} 




\appendix


\bsp	
\label{lastpage}
\end{document}